\documentclass[12pt,preprint]{aastex}
\usepackage{mathrsfs}
\usepackage{amsmath}
\usepackage{url}

\begin{document}
\title{YNOGK: A new public code for calculating null geodesics in the Kerr spacetime}
\author{Xiaolin Yang\altaffilmark{1,2,3}, Jiancheng Wang\altaffilmark{1,2}}
\altaffiltext{1} {National Astronomical Observatories, Yunnan Observatory, Chinese Academy
of Sciences,  Kunming 650011, China}
\altaffiltext{2} {Key Laboratory for the Structure and Evolution of Celestial Objects,
Chinese Academy of Sciences,  Kunming 650011, China}
\altaffiltext{3} {Graduate School, Chinese Academy of Sciences, Beijing, P.R. China}

\email{yangxl@ynao.ac.cn}

\begin{abstract}
Following \cite{dexagol2009}
we present a new public code for the fast calculation of null
geodesics in the Kerr spacetime. Using Weierstrass' and Jacobi's elliptic functions, we express all coordinates and affine parameters
as analytical and numerical functions of a parameter $p$, which is an integral value along
the geodesic. This is a main difference of our code compares with previous similar ones.
The advantage of this treatment is that the information about the turning points do not need to be specified in advance
by the user,
and many applications such as imaging, the calculation of line profiles or the observer-emitter problem, etc become root
finding problems. All elliptic integrations are computed by Carlson's elliptic integral method as
\cite{dexagol2009} did, which guarantees the fast computational speed of our code.
The formulae to compute the constants of motion given by \cite{cunnbard1973} have been extended, which
allow one readily to handle the situations, in which the emitter or the observer has arbitrary distance and motion
state with respect to the central compact object. The validation of the code has been extensively tested by its
application to toy problems from the literature. The source FORTRAN code is freely available for download on the
web.\footnote{\url{http://www1.ynao.ac.cn/~yangxl/yxl.html}}
\end{abstract}


\section{Introduction}
There are wide interests in calculating the radiative transfer near the
compact objects, such as black hole, neutron star and white dwarf.
The radiation will be affected by various effects, such as, light bending or focusing, time dilation,
Doppler boosting and gravitational redshift, under the strong gravitational field of the compact objects.
The consideration of these effects not only help us to understand the observed results,
therefor to study these compact objects, but also even
to test the correctness of the general relativity under strong gravity.
A good example is the study of the fluorescent iron line in the X-ray band at 6.4-6.9 keV,
which is seen in many active galactic nuclei especially for Seyfert galaxies
\citep{laor,fabian2000,muller,MiFa,Miniet}. The line appears broadened and skewed as a result of
the Doppler effect and gravitational redshift, thus it is an important diagnostic to
study the geometry and other properties of the accretion flow at the
vicinity of the central black hole \citep{fabian2000}. Another example is the study of SMBH
in the center of our galaxy.
It has been comprehensively accepted that in the center of our galaxy a
super-massive black hole with $\sim$ 4$\times 10^6$ $M_{\odot}$ exists
\citep{schodel2003,gebhardt2000,hopkins2008} and its shadow may be observed
directly in the radio band in the near future.
Based on the general relativistic numerical simulations of the accretion
flow, \citet{noble2007} present the first dynamically self-consistent
models of millimeter and sub-millimeter emission from Sgr $A^*$.
\citet{Yuan2009} calculated the observed images of Sgr $A^*$ with a fully general relativistic
radiative inefficient accretion flow.

A natural way to include all of the gravitational effects is to track the ray following its
null geodesic orbit. Which requires the fast and accurate computations of
the trajectory of a photon in the Kerr spacetime.
\citet{fanton1997} proposed a fast code to calculate the
accreting lines and thin disk images. \citet{cadez1998} translated
all integrations into the Legendre's standard elliptic integrals and
wrote a fast numerical code. \citet{dexagol2009} presented a new fast
public code, named geokerr, for the computing of all coordinates of null geodesics in the Kerr spacetime by
using the Carlson's elliptic integrals semi-analytically for the first time.
There are two computational methods often used in these codes, they are:
(1) the elliptic function method, which relies on the integrability of the geodesics
\citep{cunnbard1973,cunningham1975,rauchblandford,speith1995,fanton1997,cadez1998,li2005,wu2007,dexagol2009,Yuan2009},
and (2) the direct geodesic integration method 
\citep{fuerstwu2004,schnittman2006,Dolence2009,anderson2010,vincent2011,Younsi2012}.
Usually people regard the direct geodesic integration method to be a better choice than the elliptic function method,
for the direct geodesic integration method can deal with any three-dimensional accretion flow \citep{Younsi2012},
especially in radiative transfer problems which require the calculations of many points along each geodesic,
the direct integration method is simpler and faster \citep{Dolence2009}. While the the elliptic function
method is considered to be just efficient for the calculation of the emissions come from an optically thick, geometrical
thin and axisymmetric accretion disk system. But we think that the elliptic function
method based on \cite{dexagol2009} after some extensions can overcome these shortages and
not only handle any three-dimensional accretion flows readily,
but also be more efficient, flexible, and accurate, because it can compute arbitrary points on
arbitrary sections for any geodesics. The speed of the code based on this approach is still very fast for many potential
applications. While the direct geodesic integration
method must integrate the geodesic from the initial position to the interested points, the waste
of computational time is inevitable.

We present a new public code for the computation of null geodesics in the Kerr spacetime following the work of \cite{dexagol2009}.
In our code all coordinates and the affine parameters are expressed as functions of a parameter $p$, which corresponds to
$I_u$ or $I_\mu$ in \cite{dexagol2009}. Using parameter $p$ as the independent
variable, the computations are easier and simpler, thus
more convenient, mainly due to the fact that the information about the turning points does not need to be
prescribed in advance comparing with \cite{dexagol2009}. Meanwhile \cite{Yuan2009} have demonstrated
that the parameter $p$ can replace the affine parameter to be the independent variable
in radiative transfer problems. We not only take this replacing, but also used it to handle more sophisticated
applications. We extend the formulae of computing constants of motion from initial conditions to
a more general form, which can readily handle the cases in which the emitter or the observer
has arbitrary motion state and distance with respect to the central black hole.
We reduce the elliptic integrals from the motion equations derived from the Hamilton-Jacobi
equation \citep{carter68} to the Weierstrass' elliptic integrals
rather than to the Legendre's ones, because the former ones are much easier to handle. Then
we calculate these integrals by Carlson's method.

The paper is organized as follows. In section \ref{geoequ} we give
the motion equations for null geodesics. In section \ref{weifun} we
express all coordinates and affine parameters as functions
of a parameter $p$ analytically and numerically. In section \ref{imppar} we give the extended formulae
for the computation of constants of motion.  A brief introduction and discussion about the code are
given in section \ref{genmet}. In section \ref{protest} we demonstrate the testing results of
our code for toy problems in the literature. The conclusions and discussions are finally
presented in section \ref{discconc}. The general relativity calculations follow the
notational conventions of the text given by \cite{mtw}. The natural unit are used through out this
paper, in which constants G=c=1,
and the mass of the central black hole M is also taken to be 1.

\section{motion equations} \label{geoequ}
Following \citet{bardeen1972}, we
write the Kerr line element in the Boyer-Lindquist (B-L) coordinates $(t, r, \theta, \phi)$ as
\begin{equation}
 ds^2=-e^{2\nu}dt^2+e^{2\psi}(d\phi-\omega
dt)^2+e^{2\mu_1}dr^2+e^{2\mu_2}d\theta^2,
\end{equation}
where
\begin{equation}
\notag e^{2\nu}=\frac{\Sigma\Delta}{A}, \quad e^{2\psi}=\frac{\sin^2\theta A}{\Sigma}, \quad
e^{2\mu_1}=\frac{\Sigma}{\Delta}, \quad e^{2\mu_2}=\Sigma, \quad \omega=\frac{2ar}{A},
\end{equation}
\begin{equation}
\Delta=r^2-2r+a^2,\quad \Sigma=r^2+a^2\cos^2\theta, \quad
A=(r^2+a^2)^2-\Delta a^2\sin^2\theta,
\end{equation}
and $a$ is the spin parameter of the black hole.

The geodesic equations for a freely test particle read
\begin{equation}
\frac{d^2x^\alpha}{d\sigma^2}+\Gamma_{\mu\nu}^{\alpha}u^\mu u^\nu=0,
\end{equation}
where $\sigma$ is the proper time for particles and an affine parameter for photons, $u^{\nu}$ is the
four-velocity, $\Gamma_{\mu\nu}^{\alpha}$ is the connection coefficients.
\citet{carter68} found these equations are integrable under Kerr spacetime and got the differential and integral
forms of motion equations for particles by using the Hamilton-Jacobi equation.
For a photon, whose rest mass $m$ is zero, the equations of motion have the following forms:
\begin{eqnarray}
\label{defr}\Sigma \frac{dr}{d\sigma}&=&\pm\sqrt{R},\\
\notag\\
\label{deftheta}\Sigma \frac{d\theta}{d\sigma}&=&\pm\sqrt{\Theta_\theta},\\
\notag\\
\label{defphi}\Sigma\frac{d\phi}{d\sigma}&=&-(a-\frac{\lambda}{\sin^2\theta})+a\frac{T}{\Delta},\\
\notag\\
\label{deft}\Sigma\frac{dt}{d\sigma}&=&-a(a\sin^2\theta-\lambda)+(r^2+a^2)\frac{T}{\Delta},
\end{eqnarray}
\begin{eqnarray}
\label{pdefine}&&\pm\int^r\frac{dr}{\sqrt{R}}=\pm\int^\theta\frac{d\theta}{\sqrt{\Theta_\theta}},\\
\notag\\
\label{sigmadef}&&\sigma=\int^r\frac{r^2}{\sqrt{R}}dr+a^2\int^\theta\frac{\cos^2\theta}{\sqrt{\Theta_\theta}}d\theta,\\
\notag\\
\label{tdefine}&&t=\sigma+2\int^r\frac{rT}{\Delta\sqrt{R}}dr,\\
\notag\\
\label{phidefine}&&\phi=a\int^r\frac{T}{\Delta\sqrt{R}}dr+\int^\theta\frac{\lambda\csc^2\theta-a}{\sqrt{\Theta_\theta}}d\theta,
\end{eqnarray}
where
\begin{equation}\label{Rr}
R=r^4-(q+\lambda^2-a^2)r^2+2[q+(\lambda-a)^2]r-a^2q,
\end{equation}
\begin{equation}
\label{theta}\Theta_\theta=q+a^2\cos^2\theta-\lambda^2\cot^2\theta,
\end{equation}
\begin{equation}
T=r^2+a^2-\lambda a,
\end{equation}
q and $\lambda$ are constants of motion defined by
\begin{equation}
q=\frac{Q}{E^2}, \quad \lambda=\frac{L_z}{E},
\end{equation}
where $Q$ is the Carter constant,  $L_z$ is the angular momentum of the photon
about the spin axis of the black hole, $E$ is the energy measured by an observer at infinity.
The four momentum of a photon can be expressed as
\begin{eqnarray}
 p_\mu=E(-1,\pm\frac{\sqrt{R}}{\Delta},\pm\sqrt{\Theta_\theta},\lambda),
\end{eqnarray}
which is often used in the discussion of the motion of a photon.

From the equation (\ref{theta}) we know that if $q=0$ and $\theta\equiv\pi/2$, then $\Theta_\theta\equiv0$.
$\theta\equiv\pi/2$ means the motion of the photon is confined in the equatorial
plane forever \citep{chandrasekhar83}. Thus the motion equations with integral forms now
become invalid for $\Theta_\theta=0$ appeared in
the denominator. We need the motion equations with differential forms.

From equation (\ref{defr}) we have
\begin{equation}
\label{sigmadefpm}\sigma_{\mathrm{pm}} = \int^r\frac{r^2}{\sqrt{R}}dr,
\end{equation}
where the subscript $\mathrm{pm}$ means 'plane motion'. Dividing equation (\ref{defphi}) by (\ref{defr}) and integrating both sides,
we obtain
\begin{equation}
\label{phidefine2}\phi_{\mathrm{pm}}=\int^r\frac{\lambda-a}{\sqrt{R}}dr+a\int^r\frac{T}{\Delta}\frac{dr}{\sqrt{R}}.
\end{equation}
Similarly from equation (\ref{deft}) and (\ref{defr}), we obtain
\begin{eqnarray}
\notag\label{tdefine2}t_{\mathrm{pm}}&=&\int^r\frac{r^2}{\sqrt{R}}dr+2\int^r\frac{rT}{\Delta}\frac{dr}{\sqrt{R}},\\
&=&\sigma_{\mathrm{pm}}+2\int^r\frac{rT}{\Delta}\frac{dr}{\sqrt{R}}.
\end{eqnarray}

The spherical motion is an another special case, in which the photon is confined on a sphere and the motion of which can
be described by equations:
$R\equiv0$ and $dR/dr\equiv0$ \citep{bardeen1972,shakura1987}. Thus the motion equations with
integral forms also become invalid due to $R=0$ appears in the denominator. Similarly from equation (\ref{deftheta})
we have
\begin{eqnarray}
 \sigma_{\mathrm{sm}} = r^2\int^\theta\frac{d\theta}{\sqrt{\Theta_\theta}}+\int^\theta\frac{a^2\cos^2\theta}{\sqrt{\Theta_\theta}}d\theta,
\end{eqnarray}
where the subscript sm means "spherical motion". Dividing equation (\ref{defphi}) by (\ref{deftheta}) and integrating both sides,
we have
\begin{eqnarray}
 \phi_{\mathrm{sm}}=a\frac{T}{\Delta}\int^\theta\frac{d\theta}{\sqrt{\Theta_\theta}}+\int^\theta\frac{\lambda\csc^2\theta-a}{\sqrt{\Theta_\theta}}d\theta.
\end{eqnarray}
From the equations (\ref{deftheta}) and (\ref{deft}) we have
\begin{eqnarray}
 \label{tdefcm}t_{\mathrm{sm}}= \sigma_{\mathrm{sm}}+2r\frac{T}{\Delta}\int^\theta\frac{d\theta}{\sqrt{\Theta_\theta}}.
\end{eqnarray}

From equation (\ref{pdefine}), we introduce a new parameter $p$ with following definition to describe the
motion of a photon along its geodesic \citep{Yuan2009}
\begin{equation}
\label{pdefine1}p=\pm\int^r\frac{dr}{\sqrt{R}}=\pm\int^\theta\frac{d\theta}{\sqrt{\Theta_\theta}}.
\end{equation}
Because the sign ahead the integral is the same with $dr$ and $d\theta$, $p$ is always nonnegative and increases
monotonically as the photon movies along the geodesic. From the above definition, we know
that $r$ and $\theta$ are functions of $p$. In the next section, we will give the explicit forms of
these functions by using Weierstrass' and Jacobi's elliptic functions.

\section{The expressions of all coordinates as functions of $p$} \label{weifun}
\subsection{Turning points}
From the equations of motion we know that both $R$ and $\Theta_\theta$ must be nonnegative. This restriction
divides the coordinate space into allowed (where $R\geq0$ and $\Theta_\theta\geq0$) and forbidden
(where $R<0$ or $\Theta_\theta<0$) regions for the motion of a photon. The boundary points
of these regions are the so called turning points, their coordinates $r_{tp}$ and $\theta_{tp}$ satisfy equations
$R(r_{tp})=0$ and $\Theta_\theta(\theta_{tp})=0$. For a photon emitted at $r_{ini}$ and $\theta_{ini}$,
its motion will be confined between two turning points $r_{tp_1}$ and $r_{tp_2}$ for radial coordinate,
$\theta_{tp_1}$ and $\theta_{tp_2}$ for poloidal coordinate. If we assume $r_{tp_1}\leq r_{tp_2}$, and
$\theta_{tp_1}\leq\theta_{tp_2}$, then we have $r_{ini}\in[r_{tp_1}, r_{tp_2}]$ and $\theta_{ini}\in [\theta_{tp_1}, \theta_{tp_2}]$.
Because $p_r=\pm\sqrt{R}/\Delta$ and $p_\theta=\pm\sqrt{\Theta_\theta}$, if $p_r=0$ (or $p_\theta=0$)
at the initial position, we have $R(r_{ini})=0$ (or $\Theta_\theta(\theta_{ini})=0$),
therefore $r_{ini}$ (or $\theta_{ini}$) must be a turning point and equal to one of $r_{tp_1}$ and $r_{tp_2}$
(or $\theta_{tp_1}$ and $\theta_{tp_2}$).

The radial motion of a photon can be unbounded, meaning that the photon can go to the infinity
or fall into the black hole. These cases usually correspond to equation $R(r)=0$ has no real roots or
$r_{tp_1}$ is less or equal to the radius of the event horizon.
We regard the infinity and the event horizon of the black hole as two special
turning points in the radial motion, a photon will asymptotically approach them but never return from them.
Thus $r_{tp_2}$ can be the infinity and $r_{tp_1}$ can be less or equal to
$r_{h}$ ($r_{h}$ is the radius of the event horizon).

For the poloidal motion, there is also two special positions, $\theta=0$ and $\theta=\pi$, i.e., the spin axis of the black hole.
A photon with $\lambda=0$ will go through the spin axis due to zero angular momentum, and
will change the sign of its angular velocity $d\theta/d\sigma$ instantaneously,
and its azimuthal coordinate will jump from $\phi$ to $\phi\pm\pi$ \citep{shakura1987}, implying that
the spin axis is not a turning position. From the equation (\ref{theta}) we also
know that $0$ and $\pi$ are not the roots of equation $\Theta_\theta(\theta)=0$.
\subsection{$\mu$ coordinate}
\label{mucases}
Firstly, we use a new variable $\mu$ to replace
$\cos\theta$, and the equation (\ref{pdefine1}) can be rewritten as:
\begin{equation}
\label{pdefine2}p=\pm\int^r\frac{dr}{\sqrt{R}}=\pm\int^\mu\frac{d\mu}{\sqrt{\Theta_\mu}},
\end{equation}
where
\begin{equation}
\Theta_\mu=q-(q+\lambda^2-a^2)\mu^2-a^2\mu^4.
\end{equation}

Both $R$ and $\Theta_\mu$ are quartic, but the polynomial of Weierstrass' standard elliptic integral is
cubic. We need a variable transformation to make $R$ and $\Theta_\mu$ to be cubic.
We define the following constants for poloidal motion:
\begin{eqnarray}
&&b_0 = -4a^2\mu_{tp_1}^3-2(q+\lambda^2-a^2)\mu_{tp_1},\\
&&b_1 = -2a^2\mu_{tp_1}^2-\frac{1}{3}(q+\lambda^2-a^2) ,\\
&&b_2 = -\frac{4}{3}a^2\mu_{tp_1},\\
&&b_3 = -a^2,
\end{eqnarray}
where $\mu_{tp_1}=\cos\theta_{tp_1}$ and introduce a new variable $t$,
\begin{equation}
\label{tofmu}t=\frac{b_0}{4}\frac{1}{(\mu-\mu_{tp_1})}+\frac{b_1}{4}.
\end{equation}
Making transformation from $\mu$ to t, the $\mu$ part of equation (\ref{pdefine2}) can be reduced to
\begin{equation}
\label{pdef_mu}p = \pm\int^{t(\mu)}\frac{dt}{\sqrt{4t^3-g_2t-g_3}},
\end{equation}
where $g_2 = \frac{3}{4}(b_1^2-b_0b_2)$, $g_3 = \frac{1}{16}(3b_0b_1b_2-2b_1^3-b^2_0b_3).$
Using the definition of Weierstrass' elliptic function $\wp(z;g_2,g_3)$ \citep{abramstegun}, from equation (\ref{pdef_mu}),
we have $t=\wp(p\pm\Pi_\mu;g_2,g_3)$. Solving equation (\ref{tofmu}) for $\mu$,
we can express $\mu$ as the function of $p$:
\begin{equation}
\label{mup}\mu(p)=\frac{b_0}{4\wp(p\pm\Pi_\mu;g_2,g_3)-b_1}+\mu_{tp_1},
\end{equation}
where
\begin{equation}
\Pi_\mu=|\wp^{-1}[t(\mu_{ini});g_2,g_3]|.
\end{equation}
The sign ahead $\Pi_\mu$ depends on the initial value of $p_\theta$, which is
the $\theta$ component of four momentum of a photon, and
\begin{eqnarray}
\label{sign_theta}\left\{
\begin{array}{cc}
p_\theta>0,&\quad\quad\quad\quad\quad\quad\quad\quad\quad\quad\quad\quad\quad\quad+,\\
p_\theta=0,&\left\{
  \begin{array}{ccccc}
               \theta_{ini} = \theta_{tp_1},& \Pi_\mu=&n\omega&,&+,-,\\
               \theta_{ini} = \theta_{tp_2},& \Pi_\mu=&(\frac{1}{2}+n)\omega&,&+,-,\\
  \end{array}
  \right. \\
p_\theta<0,&\quad\quad\quad\quad\quad\quad\quad\quad\quad\quad\quad\quad\quad\quad-,
\end{array}
\right.
\end{eqnarray}
where $\omega$ is the period of $\wp(z;g_2,g_3)$ and $n=0,1,2,\cdots$. The sign ahead $\Pi_\mu$ can be
"$+$" or "$-$" when $p_\theta = 0$.

From the above discussion, we know that one root of equation $\Theta_\mu = 0$ is needed in the variable
transformation, namely $\mu_{tp_1}$. To avoid the complexity caused by introducing complex, we always use the real one. Luckily, equation
$\Theta_\mu=0$ always has real roots, but which is not true for equation $R(r) = 0$. For cases in which
equation $R(r)=0$ has no real roots
we will use the Jacobi's elliptic functions $\mathrm{sn}(z|k^2),\mathrm{cn}(z|k^2)$ to express $r$.

\subsection{$r$ coordinate}
If equation $R(r) = 0$ has real roots, then $r_{tp_1}$ exists, we can define the following constants by using
$r_{tp_1}$:
\begin{eqnarray}
&&b_0 =  4r_{tp_1}^3-2(q+\lambda^2-a^2)r_{tp_1}+2[q+(\lambda-a)^2],\\
&&b_1 =  2r_{tp_1}^2-\frac{1}{3}(q+\lambda^2-a^2) ,\\
&&b_2 =  \frac{4}{3}r_{tp_1},\\
&&b_3 =  1,
\end{eqnarray}
and introduce a new variable $t$,
\begin{equation}
\label{tofr}t=\frac{b_0}{4}\frac{1}{(r-r_{tp_1})}+\frac{b_1}{4}.
\end{equation}
It is similar with $\mu$, using $t$ as the independent variable, we can reduce the $r$ part of equation (\ref{pdefine2}) into the standard form of
Weierstrass' elliptical integral
\begin{equation}
\label{ptr}p = \pm\int^{t(r)}\frac{dt}{\sqrt{4t^3-g_2t-g_3}},
\end{equation}
where $g_2 = \frac{3}{4}(b_1^2-b_0b_2)$, $g_3 = \frac{1}{16}(3b_0b_1b_2-2b_1^3-b^2_0b_3).$
Taking the inverse of above equation, we get $t=\wp(p\pm\Pi_r;g_2,g_3)$. Solving equation (\ref{tofr}) for
r, we have
\begin{equation}
\label{wrp}r(p)=\frac{b_0}{4\wp(p\pm\Pi_r;g_2,g_3)-b_1}+r_{tp_1},
\end{equation}
where
\begin{equation}
\Pi_r=|\wp^{-1}[t(r_{ini});g_2,g_3]|.\quad
\end{equation}
The sign ahead $\Pi_r$ also depends on the initial value of $p_r$, which is
the $r$ component of four momentum of a photon, and
\begin{eqnarray}
\label{sign_r}\left\{
\begin{array}{cc}
p_r>0,&\quad\quad\quad\quad\quad\quad\quad\quad\quad\quad\quad\quad\quad\quad+,\\
p_r=0,&\left\{
  \begin{array}{ccccc}
               r_{ini} = r_{tp_1},& \Pi_r=&n\omega&,&+,-,\\
               r_{ini} = r_{tp_2},& \Pi_r=&(\frac{1}{2}+n)\omega&,&+,-,\\
  \end{array}
  \right. \\
p_r<0,&\quad\quad\quad\quad\quad\quad\quad\quad\quad\quad\quad\quad\quad\quad-,
\end{array}
\right.
\end{eqnarray}
where $\omega$ is the period of $\wp(z;g_2,g_3)$ and $n=0,1,2,\cdots$. The sign ahead $\Pi_r$ can be
"$+$" or "$-$" when $p_r = 0$.

If equation $R(r)=0$ has no real roots, we use the Jacobi's elliptic functions to express $r$.
Since the coefficient of $r^3$ is zero, the roots of equation $R(r)=0$ satisfy $r_1+r_2+r_3+r_4=0$.
Therefore the roots $r_1, r_2, r_3, r_4$ can be written as
\begin{eqnarray}
\notag&&r_1=u-iw, \,\,\, r_2=u+iw,\\
&&r_3=-u-iv, r_4=-u+iv.
\end{eqnarray}
Introducing two constants $\lambda_1$ and $\lambda_2$
\begin{equation}
\lambda_{1,2}=\frac{1}{2w^2}[4u^2+v^2+w^2\pm\sqrt{(4u^2+w^2+v^2)^2-4w^2v^2}],
\end{equation}
which satisfy $\lambda_1>1>\lambda_2>0$, and a new variable $t$,
\begin{equation}
\label{tofr2}t=\sqrt{\frac{\lambda_1-1}{(\lambda_1-\lambda_2)[(r-u)^2+w^2]}}\left(r-u\frac{\lambda_1+1}{\lambda_1-1}\right),
\end{equation}
we can reduce the $r$ part of equation (\ref{pdefine2}) to the Legendre's standard elliptic integral
\begin{equation}
\label{plegendre}p=\int^t\frac{dt}{w\sqrt{\lambda_1}\sqrt{(1-t^2)(1-m_2t^2)}},
\end{equation}
where
\begin{equation}
m_2=\frac{\lambda_1-\lambda_2}{\lambda_1}.
\end{equation}
Using the definition of Jacobi's elliptic function $\mathrm{sn}(z|k^2)$ \citep{abramstegun}, from equation (\ref{plegendre}) we
obtain $t=\mathrm{sn}(pw\sqrt{\lambda_1}\pm\Pi_0|m_2)$. Solving the equation (\ref{tofr2}) for $r$, we get the expression of
$r$ as the function of $p$
\begin{equation}
\label{lrp}r_{\pm}(p)=u+\frac{-2u\pm w(\lambda_1-\lambda_2)\mathrm{sn}(pw\sqrt{\lambda_1}\pm\Pi_0|m_2)|\mathrm{cn}(pw\sqrt{\lambda_1}\pm\Pi_0|m_2)|}
{(\lambda_1-\lambda_2)\mathrm{sn}^2(pw\sqrt{\lambda_1}\pm\Pi_0|m_2)-(\lambda_1-1)},
\end{equation}
where
\begin{equation}
\Pi_0=|sn^{-1}\left[t(r_{ini})|m_2\right]|.
\end{equation}
When the initial value of $p_r>0$, we have $r = r_-$, and when $p_r<0$, we have $r = r_+$. And $p_r$ can not be
zero, otherwise the initial radial coordinate $r_{ini}$ of the photon will be one root of equation $R(r) = 0$,
which is the case that has been discussed above.

\subsection{$t$ and  $\phi$ coordinates and affine parameter $\sigma$}

In this section, we will express the coordinates $t$, $\phi$ and the affine parameter $\sigma$ as the
numerical functions of the parameter $p$. All of these variables have been
expressed as the integrals of $r$ and $\theta$ in the equations (\ref{sigmadef})-(\ref{phidefine}) and
(\ref{sigmadefpm})-(\ref{tdefcm}).
The goal is achieved if we can compute all of these integrals along a geodesic for a
specified $p$. {\bf{Making transformations}} from $r$ and $\mu$ to a new variable $t$
(defined by equations (\ref{tofmu}), (\ref{tofr}) and (\ref{tofr2})),
we will compute these integrals under the new variable $t$. For simplicity we use
$F_r(t) \mbox{ and } F_\theta(t)$ to denote the
complicated integrands (see below) for $r \mbox{ and } \theta$ respectively.

\begin{figure}[ht!]
\begin{center}
\includegraphics[width=0.45\textwidth]{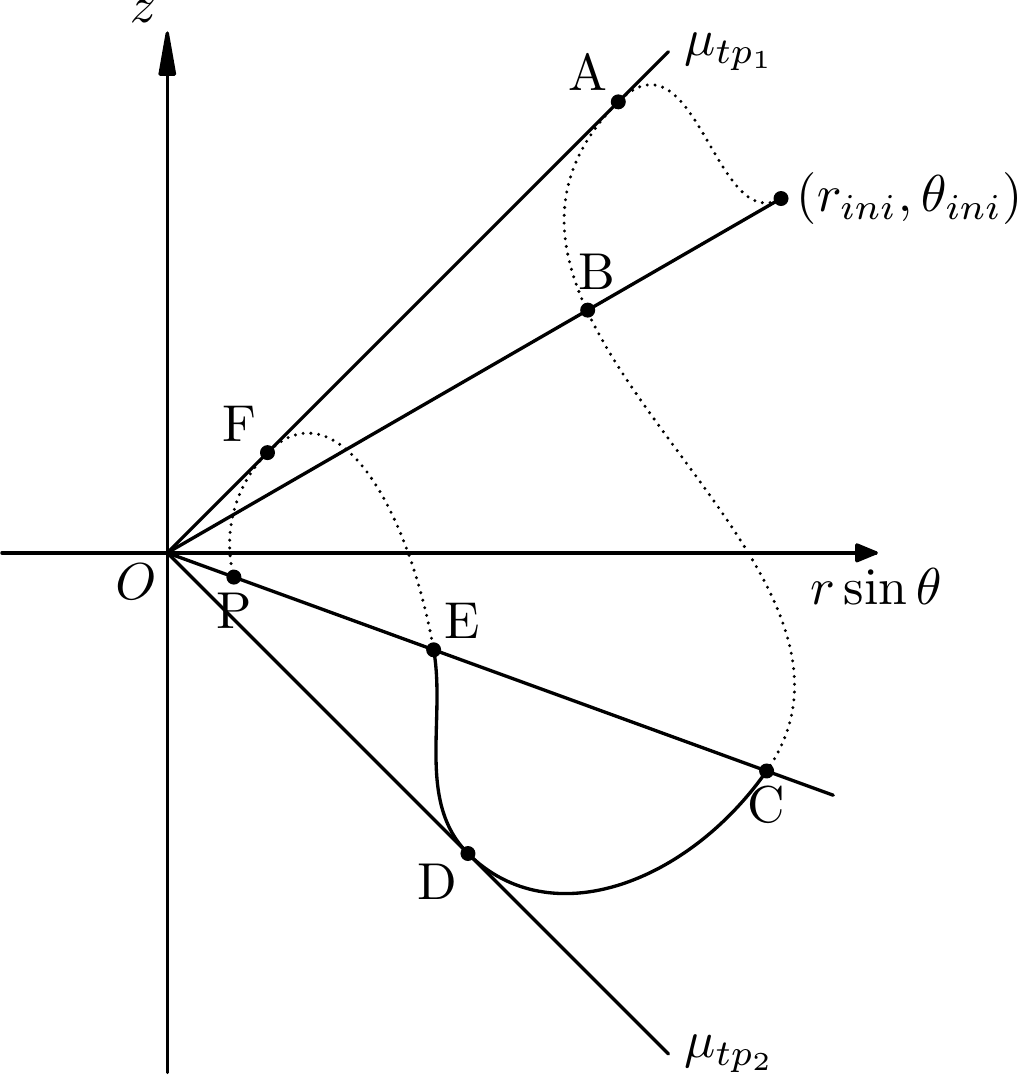}
\caption{\label{tpmu}This figure illustrates the motion of a photon in
the $\theta$ coordinate, which has been projected onto the $r-\theta$ plane.
The motion is confined between two turning points $\mu_{p1}$ and
$\mu_{p2}$. A, D and F indicate the positions where the
photon reaches the turning points and P indicates the position of the photon.
The path between any two neighboring turning points (such as DA, DF)
has the maximum monotonic length and the integrals of $\theta$ should be evaluated
along each monotonic section and summed. The doted (such as CA, EF and PF) and solid
(such as DC and DE) lines represent the integral paths of $I_1$ and
$I_2$ (see text) respectively. Obviously the BC section is the integral
path of $I_0$, and one has $I_0=\int^C_B$,
$I_1=\int^A_C=\int^F_E=\int^F_P$, $I_2=\int^C_D=\int^E_D$, etc.}
\end{center}
\end{figure}

Firstly, we discuss the integral path, which starts from the initial position and terminates at the photon.
If the photon encounters turning points along the geodesic, then the whole integral path is
not monotonic, as shown in Figure.\ref{tpmu} for poloidal motion (radial motion is similar). In this figure the projected poloidal
motion of a photon onto the $r$-$\theta$
plane is illustrated. The motion is confined between two turning points: $\mu_{tp_1}$ and $\mu_{tp_2}$.
The photon encounters the turning points for three times. Obviously any sections of the path which
contain one or more than one turning points is not monotonic, such as path CDE, EFP etc. The path between
any two neighboring turning points has the maximum monotonic length and the total integrals should be computed
along each of them and summed.

There are four important points involved in the limits of these integrals, i.e., $\mu_{tp_1}$, $\mu_{tp_2}$, $\mu_{ini}$ and $\mu_p$,
they are the $\mu$ coordinates of turning points, initial point and the photon position for a given $p$ respectively.
And the values of these points corresponding to the new variable $t$ are $t_{tp_1}$, $t_{tp_2}$, $t_{ini}$,
which can be calculated from equation (\ref{tofmu}),
and $t_p$, which can be calculated from $t=\wp(p\pm\Pi_\mu;g_2,g_3)$ with a given $p$. Because the function $t=t(\mu)$ expressed by equation
(\ref{tofmu}) is monotonically increasing,
we have $t_{tp_2}\leq t_{ini}\leq t_{tp_1}$ and $t_{tp_2}\leq t_{p}\leq t_{tp_1}$.

If we use $Nt_1$ and $Nt_2$ to denote
the number of times of a photon meeting the turning points $\mu_{tp_1}$ and $\mu_{tp_2}$ for a given $p$ respectively,
and define the following integrals (cf. Figure \ref{tpmu}):
\begin{eqnarray}
\notag&&I_0=\int^{t_p}_{t_{ini}}F_\theta(t) dt,\quad I_1=\int^{t_{tp_1}}_{t_{p}}F_\theta(t) dt,\quad I_2=\int^{t_p}_{t_{tp_2}}F_\theta(t) dt,\\
&&I_{01}=I_0+I_1=\int^{t_{tp_1}}_{t_{ini}}F_\theta(t) dt,\quad I_{02}=I_2-I_0=\int^{t_{ini}}_{t_{tp_2}}F_\theta(t) dt,
\end{eqnarray}
then the integrals of $\theta$ in $\sigma$, $t$ and $\phi$ then can be written as (cf. Figure \ref{tpmu})
\begin{eqnarray}
\notag\label{integralmu}\sigma_\theta\,\,\,(\,\,\phi_\theta,\,\,\,t_\theta\,\,)&=&
-\mathrm{sign}(p_\theta)I_0+2Nt_1I_1+2Nt_2I_2,\\
&=&-[\mathrm{sign}(p_\theta)+2Nt_1-2Nt_2]I_0+2Nt_1I_{01}+2Nt_2I_{02},
\end{eqnarray}
where $p_\theta$ is the $\theta$ component of the initial four momentum of a photon.
In order to evaluate the above expression, we need to know $Nt_1$ and $Nt_2$ for a given $p$.
Similarly if we define five integrals from the equation (\ref{pdef_mu}) as follows:
\begin{eqnarray}
&&\notag\label{ppdef}p_0=\int^{t_p}_{t_{\mathrm{ini}}}\frac{dt}{\sqrt{W(t)}},\quad
p_1=\int^{t_{tp_1}}_{t_p}\frac{dt}{\sqrt{W(t)}},\quad
p_2=\int^{t_p}_{t_{tp_2}}\frac{dt}{\sqrt{W(t)}}, \\
&& p_{01}=p_0+p_1=\int^{t_{tp_1}}_{t_{\mathrm{ini}}}\frac{dt}{\sqrt{W(t)}},\quad
p_{02} = p_2-p_0=\int^{t_{\mathrm{ini}}}_{t_{tp_2}}\frac{dt}{\sqrt{W(t)}},
\end{eqnarray}
where $W(t)=4t^3-g_2t-g_3$, and we will get the following identity:
\begin{eqnarray}
\notag\label{poft1t2}p&=&-\mathrm{sign}(p_\theta)p_0+2Nt_1p_1+2Nt_2p_2,\\
&=&-[\mathrm{sign}(p_\theta)+2Nt_1-2Nt_2]p_0+2Nt_1p_{01}+2Nt_2p_{02}.
\end{eqnarray}
And notice that $Nt_1$ and $Nt_2$ are not arbitrary and related to the initial
direction of the photon in poloidal motion. For $p_\theta>0$ (or $p_\theta=0$ and $\theta_{ini}=\theta_{tp_1}$), they will increase as
\begin{eqnarray*}
&&Nt_1= 0\quad0\quad1\quad1\quad2\quad2\cdots,\\
&&Nt_2= 0\quad1\quad1\quad2\quad2\quad3\cdots.
\end{eqnarray*}
For $p_\theta<0$ (or $p_\theta=0$ and $\theta_{ini}=\theta_{tp_2}$), they will increase as
\begin{eqnarray*}
&&Nt_1= 0\quad1\quad1\quad2\quad2\quad3\cdots,\\
&&Nt_2= 0\quad0\quad1\quad1\quad2\quad2\cdots.
\end{eqnarray*}
For a given $p$, we find that there always exists one pair of $Nt_1$ and $Nt_2$, which satisfy
equation (\ref{poft1t2}) and they are the number of a photon meeting the turning points.
With $Nt_1$ and $Nt_2$, the equation (\ref{integralmu}) now can be evaluated readily.

For $r$ coordinate, the process is similar with the above. $Nt_1$ and $Nt_2$ also represent
the number of times of a photon meeting the turning points $r_{tp_1}$ and $r_{tp_2}$ respectively.
Five integrals are defined as:
\begin{eqnarray}
&&\notag I_0=\int_{t_{ini}}^{t_p}F_r(t) dt,\quad I_1=\int^{t_{tp_1}}_{t_p}F_r(t) dt, \quad
I_2 = \int^{t_{p}}_{t_{tp_2}} F_r(t) dt,\\
&& I_{01}=I_0+I_1=\int^{t_{tp_1}}_{t_{ini}}F_r(t)dt, \quad
I_{02}=I_2-I_0=\int^{t_{ini}}_{t_{tp_2}}F_r(t)dt,
\end{eqnarray}
where $t_p = \wp(p\pm\Pi_r;g_2,g_3)$ (or $t_p=\mathrm{sn}(pw\sqrt{\lambda_1}\pm\Pi_0|m_2)$ when equation
$R(r)=0$ has no real roots), and
$t_{p_1}$, $t_{p_2}$ and $t_{ini}$ are calculated from equation (\ref{tofr}) or (\ref{tofr2}).
Then the integrals of $r$ in $\sigma, t, \phi$ can be written as
\begin{eqnarray}
\notag\label{integralr}\sigma_r\,\,\,(\,\,t_r,\,\, \phi_r\,\,)&=&-\mathrm{sign}(p_r)I_0+2Nt_1I_1+2Nt_2I_2,\\
&=&-[\mathrm{sign}(p_r)+2Nt_1-2Nt_2]I_0+2Nt_1I_{01}+2Nt_2I_{02}.
\end{eqnarray}
To get $Nt_1$ and $Nt_2$, we define $p_0$, $p_1$ and $p_2$ from equation (\ref{ptr}) as:
\begin{eqnarray}
&&\notag p_0=\int^{t_p}_{t_{ini}}\frac{dt}{\sqrt{W(t)}},\quad
p_1=\int^{t_{p_1}}_{t_p}\frac{dt}{\sqrt{W(t)}},\quad
p_2=\int^{t_{p}}_{t_{p_2}}\frac{dt}{\sqrt{W(t)}},\\
&&p_{01}=p_0+p_1=\int^{t_{tp_1}}_{t_{ini}}\frac{dt}{\sqrt{W(t)}},\quad
p_{02}=p_2-p_0=\int^{t_{ini}}_{t_{tp_2}}\frac{dt}{\sqrt{W(t)}}.
\end{eqnarray}
For a given $p$, we have
\begin{eqnarray}
\notag\label{pofr}p&=&-\mathrm{sign}(p_r)p_0+2Nt_1p_1+2Nt_2p_2,\\
&=&-[\mathrm{sign}(p_r)+2Nt_1-2Nt_2]p_0+2Nt_1p_{01}+2Nt_2p_{02}.
\end{eqnarray}
To get $Nt_1$ and $Nt_2$ from above equation one just needs to notice that
when $p_r>0$ (or $p_r=0$ and $r_{ini}=r_{tp_1}$) they will
increase as
\begin{eqnarray*}
&&Nt_1= 0\quad0\quad1\quad1\quad2\quad2\cdots,\\
&&Nt_2= 0\quad1\quad1\quad2\quad2\quad3\cdots,
\end{eqnarray*}
when $p_r<0$ (or $p_r=0$ and $r_{ini}=r_{tp_2}$) they will increase as
\begin{eqnarray*}
&&Nt_1= 0\quad1\quad1\quad2\quad2\quad3\cdots,\\
&&Nt_2= 0\quad0\quad1\quad1\quad2\quad2\cdots.
\end{eqnarray*}
Similarly for a given $p$, there is one pair of $Nt_1$ and $Nt_2$ satisfies equation (\ref{pofr}).
With $Nt_1$ and $Nt_2$ equation (\ref{integralr}) now can be evaluated.
In many cases the number of a photon meeting the turning points in $r$ is less
than 2, especially when $r_{tp_1}$ is less or equal to $r_h$, or $r_{tp_2}$ is
infinity, or equation $R(r)=0$ has no real roots, both $Nt_1$ and $Nt_2$ will be zero.

\subsection{Reductions to Carlson's elliptic integrals}
In previous sections, four coordinates
$r,\theta,\phi,t$ and the affine parameter $\sigma$ have been expressed as functions of $p$, and in which
many elliptic integrals need to be calculated.
In this section, we shall reduce these integrals
into standard forms and then evaluate them by Carlson's method as \citet{dexagol2009} did.

Firstly, we introduce two notations $J_k(h)$ and $I_k(h)$ with following definitions:
\begin{eqnarray}
&&J_k(h) = \int^{t_2}_{t_1}\frac{dt}{(t-h)^k\sqrt{4t^3-g_2t-g_3}},\\
&&I_k(h) = \int^{r_2}_{r_1}\frac{dr}{(r-h)^k\sqrt{(r^2-2ur+u^2+w^2)(r^2+2ur+u^2+v^2)}},
\end{eqnarray}
where $k$ is an integer. From equation (\ref{pdefine}), we get one of the standard forms as
\begin{equation}
\label{44}J_0=\int^{t_2}_{t_1}\frac{dt}{\sqrt{4t^3-g_2t-g_3}}.
\end{equation}
After being reduced to $J_0$, the forms of integrals of $r$ and $\theta$ in (\ref{pdefine}) are exactly same. Noticing the definition of
parameter $p$, we have $J_0=p$.
The radial integrals in equation (\ref{sigmadef}) are reduced to
\begin{eqnarray}
\sigma_r&=&\frac{b_0^2}{16}J_2\left(\frac{b_1}{4}\right)+\frac{b_0r_{tp_1}}{2}J_1\left(\frac{b_1}{4}\right)+r^2_{tp_1}p,
\end{eqnarray}
where $J_0$ has been replaced by $p$. The radial integrals in equation (\ref{tdefine}) can be reduced to
\begin{eqnarray}
 t_r=\sigma_r+\frac{b_0}{2}J_1\left(\frac{b_1}{4}\right)+(2r_{tp_1}+4+A_{t+}-A_{t-})p-B_{t+}J_1(t_+)+B_{t-}J_1(t_-),
\end{eqnarray}
where
\begin{eqnarray}
\notag r_\pm&=&1\pm\sqrt{1-a^2},\\
\notag A_{t\pm}&=&\frac{2[r_\pm(4-a\lambda)-2a^2]}{(r_+-r_-)(r_{tp_1}-r_{\pm})},\\
\notag B_{t\pm}&=&\frac{[r_\pm(4-a\lambda)-2a^2]b_0}{2(r_+-r_-)(r_{tp_1}-r_{\pm})^2},\\
t_\pm&=&\frac{b_1}{4}+\frac{b_0}{4(r_\pm-r_{tp_1})}.
\end{eqnarray}
Similarly the radial integrals in the equation (\ref{phidefine}) have the following form
\begin{equation}
\phi_r=a[(A_{\phi+}-A_{\phi-})p-B_{\phi+}J_1(t_+)+B_{\phi-}J_1(t_-)],
\end{equation}
where
\begin{eqnarray}
\notag A_{\phi\pm}&=&\frac{2r_\pm-a\lambda}{(r_+-r_-)(r_{tp_1}-r_\pm)},\\
B_{\phi\pm}&=&\frac{(2r_\pm-a\lambda)b_0}{4(r_+-r_-)(r_{tp_1}-r_\pm)^2}.
\end{eqnarray}
When equation $R(r)=0$ has no real roots, the integrals of $r$ in $\sigma$, $t$ and $\phi$
can be written as:
\begin{eqnarray}
&&\sigma_r = I_{-2}(0),\\
&&t_r = \sigma_r+4p+2I_{-1}(0)+C_{t+}I_{1}(r_+)-C_{t-}I_1(r_-),\\
&&\phi_r = a[C_{\phi+}I_1(r_+)-C_{\phi-}I_1(r_-)],
\end{eqnarray}
where
\begin{eqnarray}
C_{t\pm} = \frac{2[r_\pm(4-a\lambda)-2a^2]}{r_+-r_-},\quad
C_{\phi\pm} = \frac{2r_\pm-a\lambda}{r_+-r_-}.
\end{eqnarray}
The integrals concerning $\mu$ in the equation (\ref{sigmadef}) are reduced to
\begin{equation}
\sigma_\mu=a^2\left[\frac{b_0^2}{16}J_2\left(\frac{b_1}{4}\right)+\frac{b_0\mu_{tp_1}}{2}J_1\left(\frac{b_1}{4}\right)+\mu^2_{tp_1}p\right],
\end{equation}
and $t_\mu = \sigma_\mu$.
The $\mu$ integrals in the equation (\ref{phidefine}) can be reduced to
\begin{equation}
\label{48}\phi_\mu=\lambda\left[\frac{p}{1-\mu_{tp}^2}+W_{\mu_+}J_1(t_+)-W_{\mu_-}J_1(t_-)\right],
\end{equation}
where
\begin{eqnarray}
\notag W_{\mu_\pm}&=&\frac{b_0}{8(-1\pm\mu_{tp_1})^2},\\
t_\pm&=&\frac{b_1}{4}+\frac{b_0}{4(\pm1-\mu_{tp_1})}.
\end{eqnarray}
Finally we have
\begin{eqnarray}
\notag\sigma &= &\sigma_r+\sigma_\mu,\\
\notag t&=&t_r+t_\mu,\\
\phi&=&\phi_r+\phi_\mu.
\end{eqnarray}
From equations (\ref{44})-(\ref{48}) we know that the integrals need to be calculated are $J_0$, $J_1$, $J_2$,
and $I_1$, $I_{-1}$, $I_{-2}$.
Now we use the Carlson's method to evaluate them.
When equation $4t^3-g_2t-g_3$=0 has three real roots denoted
by $e_1$, $e_2$ and $e_3$,  $J_k(h)$ can be written as \citep{carlsonQUART}
\begin{eqnarray}
\notag J_k(h)&=&s_h\frac{1}{2}\int^x_y\frac{dt}{\sqrt{(t-e_1)(t-e_2)(t-e_3)(t-h)^{2k}}},\\
&=&s_h\frac{1}{2}[-1,-1,-1,-2k],
\end{eqnarray}
where $s_h$=sign$[(y-h)^k]$.
When equation $4t^3-g_2t-g_3=0$ has one real root $e_1$ and one pair of complex conjugate roots $u+iv$ and $u-iv$, $J_k(h)$ can be written as
\citep{carlson91}
\begin{eqnarray}
\notag J_k(h)&=&s_h\frac{1}{2}\int^x_y\frac{dt}{\sqrt{(t-e_1)(t^2-2ut+u^2+v^2)(t-h)^{2k}}},\\
&=&s_h\frac{1}{2}[-1,-1,-1,-2k].
\end{eqnarray}
From the equations (\ref{tofmu}) and (\ref{tofr}), we know that when $r = r_{tp_1}$ or
$\mu = \mu_{tp_1}$, $t$ will be $\infty$, thus one limit of these integrals can be infinity.

The integrals $I_{k}(h)$ can be reduced to Carlson's integrals directly \citep{carlsonDOUBLE}
\textbf{\begin{eqnarray}
\notag I_{k}(h) &=&s_h\int^x_y\frac{dr}{\sqrt{(r^2-2ur+u^2+w^2)(r^2+2ur+u^2+v^2)(r-h)^{2k}}},\\
&=&s_h[-1,-1,-1,-1,-2k],
\end{eqnarray}}
where $[p_1,\cdots,p_k]$ is a symbol used by Carlson to denote the elliptic
integrals and has the following definition:
\begin{eqnarray}
[p_1,\cdots,p_k] = \int^x_y \prod_{i = 1}^{k}(a_i+b_it)^{p_i/2}dt,
\end{eqnarray}
which can be evaluated by the formulae provided in \citet{carlsonQUART,carlson89,carlson91,carlsonDOUBLE}.

\section{Constants of motion}
\label{imppar}
\label{impactpara}
\subsection{Basic equations}
In previous sections we have expressed all coordinates as functions of a parameter $p$ and discussed
how to calculate them by Carlson's method. But before the calculation one needs to specify the constants of
motion and $p_r, p_\theta$, which determine the signs ahead $\Pi_\mu, \Pi_r, \Pi_0$
and also how the number of turning points increasing. In this section we shall discuss how to
compute $\lambda$ and $q$ and $p_r, p_\theta$ from $\bar{p}_{(a)}$, which are the components of the four-momentum measured
in the LNRF reference and have been specified by the user.

Firstly, following \cite{bardeen1972} we introduce the LNRF (locally nonrotating frame) observers or the ZAMO
(zero angular momentum observer),
the basis vectors of the orthonormal tetrad of them are given by
\begin{equation}
\label{simple}\mathbf{e}_{(a)}(\mathrm{LNRF})=e_{{(a)}}^{\nu}\partial_{\nu},
\end{equation}
where
\begin{eqnarray}
 e_{{(a)}}^{\nu}=\left(\begin{array}{cccc}
  e^{-\nu}&0 &0 & \omega e^{-\nu}\\
  0       &e^{-\mu_1} &0 & 0\\
  0&0& e^{-\mu_2} &0\\
  0 & 0 & 0 & e^{-\psi}
 \end{array}\right).
\end{eqnarray}
And the covariant components of the four momentum of a photon in the B-L coordinate can be expressed as
\begin{eqnarray}
\label{fourp}p_\mu=E\left(-1,s_r\frac{\sqrt{R}}{\Delta},s_\theta\sqrt{\Theta_{\theta}},\lambda\right),
\end{eqnarray}
where $s_r$ and $s_\theta$ are signs of r and $\theta$ components. One can easily show that
$\bar{p}_{(a)} = e_{(a)}^\mu p_\mu$, namely \citep{shakura1987}
\begin{eqnarray}
\label{energy_t}\bar{p}_{(t)}&=&-Ee^{-\nu}(1-\lambda \omega),\\
\label{energy_r}\bar{p}_{(r)}&=&s_r Ee^{-\mu_1}\frac{\sqrt{R}}{\Delta},\\
\label{energy_theta}\bar{p}_{(\theta)}&=&s_\theta Ee^{-\mu_2} \sqrt{\Theta_\theta},\\
\label{energy_phi}\bar{p}_{(\phi)}&=&E\lambda e^{-\psi}.
\end{eqnarray}
From equations (\ref{energy_r}) and (\ref{energy_theta}) we have $s_r$=$\mathrm{sign}(\bar{p}_{(r)})$ and $s_\theta$=$\mathrm{sign}(\bar{p}_{(\theta)})$,
which determine the initial direction of the photon in the B-L system (cf. equations (\ref{sign_theta}) and (\ref{sign_r})), thus
determine the way how the number of the turning points increasing.

Solving equations (\ref{energy_t}) and (\ref{energy_phi}) simultaneously for $\lambda$, one obtains
\begin{eqnarray}
\label{lambda}\lambda = \frac{\sin\theta \bar{p}_{(\phi)}/\bar{p}_{(t)}}{-\sqrt{\Delta}\Sigma/A+\omega\sin\theta \bar{p}_{(\phi)}/\bar{p}_{(t)}}.
\end{eqnarray}
Using $\lambda$ and equation (\ref{energy_t}), one obtains $E=\bar{p}_{(t)}e^\nu/(1-\lambda \omega)$.
Using $\lambda$ and $E$, from equation (\ref{energy_theta}) one obtains the formula of calculating
the motion constant $q$,
\begin{eqnarray}
\label{q}q=\left[\left(\frac{\bar{p}_{(\phi)}/\bar{p}_{(t)}}{-\sqrt{\Delta}\Sigma/A+\omega\sin\theta \bar{p}_{(\phi)}/\bar{p}_{(t)}}\right)^2-a^2\right]\cos^2\theta
+\left[\frac{\bar{p}_{(\theta)}}{\bar{p}_{(t)}}(1-\lambda \omega)\right]^2\frac{A}{\Delta}.
\end{eqnarray}
Thus we have obtained the basic equations (\ref{lambda}) and (\ref{q})
connecting $\lambda$, $q$ and the components of four momentum $\bar{p}_{(a)}$ measured in the LNRF reference.
When $\bar{p}_{(a)}$ are given the constants of motion and the initial direction of the photon are both uniquely
determined.

To prescribe $\bar{p}_{(a)}$, one should notice that they satisfy following equation
\begin{eqnarray}
-\bar{p}_{(t)}^2+\bar{p}_{(r)}^2+\bar{p}_{(\theta)}^2+\bar{p}_{(\phi)}^2=0,
\end{eqnarray}
thus there are only three independent components.
Obviously the user can specify the four momentum $\bar{p}_{(a)}$ directly in LNRF or equivalently specify
$p'_{(a)}$ in anyother reference frame of his/her own choice and then
to transform it to the LNRF reference by a Lorentz transformation, i.e., $\bar{p}_{(a)}=\alpha_{a}^{(b)}p'_{(b)}$,
where $\alpha_{a}^{(b)}$ is the transformation matrix. From equations (\ref{lambda}) and (\ref{q}) we know that
what one needs is just $\bar{p}_{(i)}/\bar{p}_{(t)}$ and
\begin{eqnarray}
\label{pivspt}\frac{\bar{p}_{(i)}}{\bar{p}_{(t)}}=\frac{\alpha_i^{(t)}+\alpha_{i}^{(j)}p'_{(j)}/p'_{(t)}}
{\alpha_t^{(t)}+\alpha_{t}^{(j)}p'_{(j)}/p'_{(t)}}.
\end{eqnarray}
The $\alpha_{a}^{(b)}$ should be specified by the user according to his/her needs.
\footnote{
When a reference frame $K'$ has physical velocities
$\upsilon_r,\upsilon_\theta,\upsilon_\phi$ with respect to a LNRF, the general Lorentz transformation matrix has six independent
parameters, i.e., $\alpha_a^{(b)}=\alpha_{a}^{(b)}(\theta_1,\theta_2,\theta_3;\upsilon_r,\upsilon_\theta,\upsilon_\phi)$,
where $\theta_i$ are the angles between the corresponding spacial basis vectors of the two references. If $\theta_i=0$, the
matrix can be written as follows \citep{mtw}
\begin{eqnarray}
\label{matrix1}\alpha^{(b)}_{a}=\left(\begin{array}{cccc}
\gamma & -\gamma\upsilon_r & -\gamma\upsilon_\theta & -\gamma\upsilon_\phi\\
-\gamma\upsilon_r & 1+\gamma^2\upsilon^2_r/(1+\gamma) & \gamma^2\upsilon_r\upsilon_\theta/(1+\gamma) & \gamma^2\upsilon_r\upsilon_\phi/(1+\gamma)\\
-\gamma\upsilon_\theta & \gamma^2\upsilon_\theta\upsilon_r/(1+\gamma) & 1+\gamma^2\upsilon_\theta^2/(1+\gamma) & \gamma^2\upsilon_\theta\upsilon_\phi/(1+\gamma)\\
-\gamma\upsilon_\phi & \gamma^2\upsilon_\phi\upsilon_r/(1+\gamma) & \gamma^2\upsilon_\phi\upsilon_\theta/(1+\gamma) & 1+\gamma^2\upsilon_\phi^2/(1+\gamma)
\end{array}\right),
\end{eqnarray}
where $\gamma=[1-(\upsilon_r^2+\upsilon_{\theta}^2+\upsilon_{\phi}^2)]^{-1/2}$,
and its inverse form
\begin{eqnarray}
\label{matrix2}\widetilde{\alpha}_{(a)}^b=\left(\begin{array}{cccc}
\gamma & \gamma\upsilon_r & \gamma\upsilon_\theta & \gamma\upsilon_\phi\\
\gamma\upsilon_r & 1+\gamma^2\upsilon^2_r/(1+\gamma) & \gamma^2\upsilon_r\upsilon_\theta/(1+\gamma) & \gamma^2\upsilon_r\upsilon_\phi/(1+\gamma)\\
\gamma\upsilon_\theta & \gamma^2\upsilon_\theta\upsilon_r/(1+\gamma) & 1+\gamma^2\upsilon_\theta^2/(1+\gamma) & \gamma^2\upsilon_\theta\upsilon_\phi/(1+\gamma)\\
\gamma\upsilon_\phi & \gamma^2\upsilon_\phi\upsilon_r/(1+\gamma) & \gamma^2\upsilon_\phi\upsilon_\theta/(1+\gamma) & 1+\gamma^2\upsilon_\phi^2/(1+\gamma)
\end{array}\right).
\end{eqnarray}
}

\begin{figure}[ht!]
\begin{center}
\includegraphics[width=0.7\textwidth]{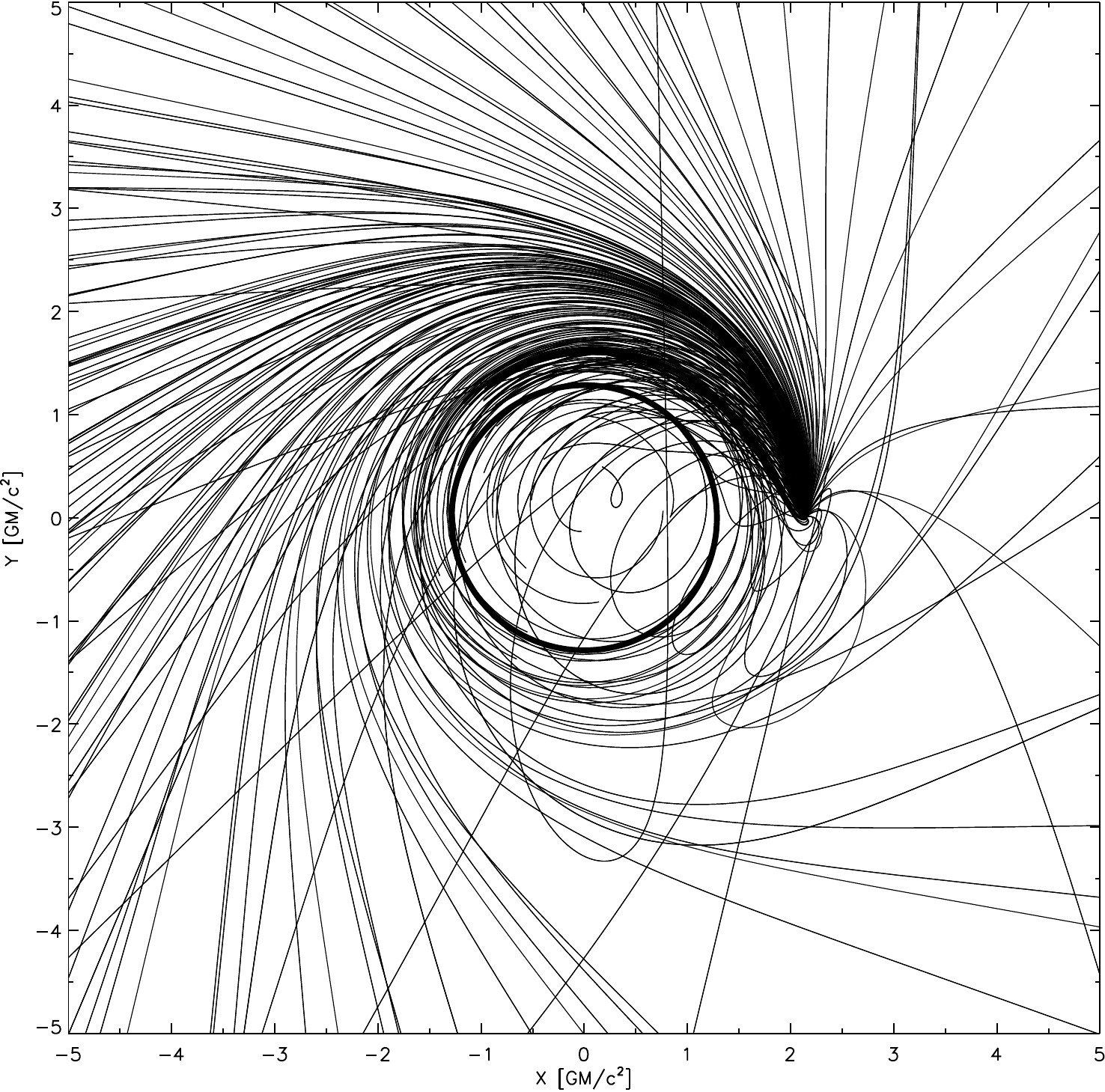}
\caption{\label{equatorialplane}A set of geodesics emitted isotropically from a
particle orbits around a black hole in the marginally stable circular orbit with $a$=0.9375. x and y
are pseudo-Cartesian coordinates in the equatorial plane of the black hole. The figure shows
the light bending and beaming effects clearly. A circle in the center represents the boundary
of the event horizon.}
\end{center}
\end{figure}

As an example, in Figure \ref{equatorialplane}, we show a group of null geodesics emitted isotropically from a particle
moving around a black hole in a marginally stable circular orbit ($r_{ms}$) with a=0.9375. And the physical velocities of the
particle with respect to the LNRF are $\upsilon_r=\upsilon_\theta=0,\upsilon_\phi=e^{\psi-\nu}(\Omega-\omega)|_{r=r_{ms}}$. The
four-momentum $p'_{(a)}$ are specified isotropically in the reference of the particle and then transformed to the LNRF by the
Lorentz transformation expressed by equation (\ref{matrix1}), i.e., $\bar{p}_{(a)}=\alpha_{a}^{(b)}p'_{(b)}$. With $\bar{p}_{(a)}$
the constants of motion are computed readily. The light bending and
beaming effects are illustrated obviously in this figure.

In the next section, we shall discuss how to compute $\bar{p}_{(i)}/\bar{p}_{(t)}$ from impact
parameters, which play a key role in imaging. And for simplicity we shall use the
transformation expressed by the equations (\ref{matrix1}) and (\ref{matrix2}) if the observer has
motion.

\subsection{Calculation of motion constants from impact parameters}
From the works of \citet{cunnbard1973} and
\citet{cunningham1975}, we know that $\lambda$ and $q$ can be
calculated from impact parameters, usually denoted by $\alpha$, $\beta$, which
are the coordinates of the hitting position of a photon on
the photographic plate of the observer.
The formulae provided by them read as follows
\citep{cunnbard1973}
\begin{eqnarray}
\label{alpha_inf}\lambda&=&-\alpha\sin\theta_{obs},\\
\label{beta_inf}q&=&\beta^2+(\alpha^2-a^2)\cos^2\theta_{obs}.
\end{eqnarray}
The above equations are valid only when the distance between the observer and the
emitter is infinite and the observer is stationary. Practically the distance
is not infinite, otherwise the integrals of coordinate $t$ will be divergent.
When the distance is finite, the above formulae
should be modified. We extend those formulae to general situations, in which
both the finite distance and the motion state of the observer are considered.

To consider the finite distance is very easy. One just needs to substitute the coordinates $r_{obs}, \theta_{obs}$
of the observer into equations (\ref{lambda}) and (\ref{q}) in the calculation. While to consider the motion state is more complicated.
Obviously we can distinguish the motion states of the observer into two kinds. In the first one
the observer is stationary and in the second one the observer has physical velocities $\upsilon_r,\upsilon_\theta,\upsilon_\phi$ with
respect to the LNRF reference.

In the first kind, the observer is just a LNRF observer and whose orthonormal tetrad is given by equation (\ref{simple}),
namely $\mathbf{e}_{(a)}(\mathrm{obs})=\mathbf{e}_{(a)}(\mathrm{LNRF})$.
While in the second kind, the tetrad of the observer can be created by a Lorentz transformation, i.e.,
$\mathbf{e}_{(a)}(\mathrm{obs})=\widetilde{\alpha}_{(a)}^b \mathbf{e}_{(b)}(\mathrm{LNRF})$.
Here $\widetilde{\alpha}_{(a)}^b$ is given by the equation (\ref{matrix2}).

\begin{figure}[ht!]
\begin{center}
\includegraphics[width=0.7\textwidth]{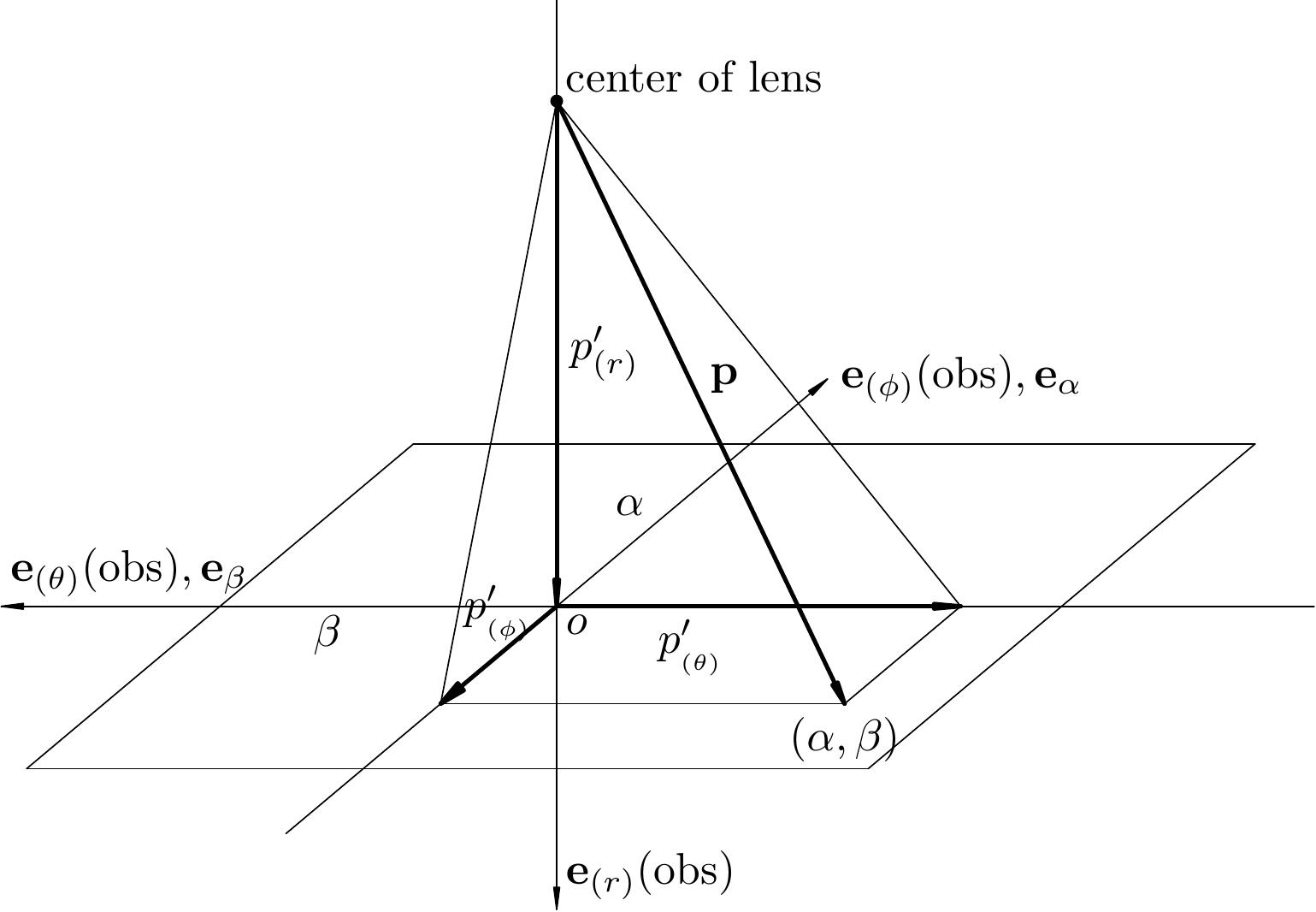}
 \caption{\label{lens}The figure shows the hitting of a photon on
the photographic plate of the observer, from which the relationships between the
impact parameters $\alpha,\beta$ and the components $p'_{(a)}$ of the four
momentum of the photon are derived.
Before hitting the plate, all
photons will go through the center of the lens.
$\mathbf{e}_{(r)}(\mathrm{obs}),\mathbf{e}_{(\theta)}(\mathrm{obs}) \mbox{ and }
\mathbf{e}_{(\phi)}(\mathrm{obs})$ are the contravariant basis vectors of the
frame, and the basis vectors of $\alpha,\beta$ coordinates $\mathbf{e}_\alpha$, $\mathbf{e}_\beta$ are
aligned with $\mathbf{e}_{(\phi)}(\mathrm{obs})$, $\mathbf{e}_{(\theta)}(\mathrm{obs})$ respectively.}
\end{center}
\end{figure}

As shown in Figure \ref{lens}, we plot the image of a photon hitting on the
photographic plate. The plate is located in the plane determined by the basis vectors
$\mathbf{e}_{(\theta)}(\mathrm{obs})$ and $\mathbf{e}_{(\phi)}(\mathrm{obs})$, and in
which an orthonormal coordinate system $\alpha$, $\beta$ has been established.
The basis vectors $\mathbf{e}_\alpha$ and $\mathbf{e}_\beta$ of the system are aligned with $\mathbf{e}_{(\phi)}(\mathrm{obs})$
and $\mathbf{e}_{(\theta)}(\mathrm{obs})$ respectively.
All photons will go through the center of the Lens before hitting on the plate.
From this figure, one obtains the relationships between the impact
parameters and $p'_{(a)}$ as follows:
\begin{eqnarray}
\label{alpha}\alpha=\left. r scal
\frac{p'_{(\phi)}}{p'_{(r)}}\right|_{r=r_{obs},\,\theta=\theta_{obs}},\\
\label{beta}\beta=\left. r scal
\frac{p'_{(\theta)}}{p'_{(r)}}\right|_{r=r_{obs},\,\theta=\theta_{obs}}.
\end{eqnarray}
And obviously one can read off
$p'_{(r)}> 0$, $\alpha p'_{(\phi)}\ge 0$ and $\beta p'_{(\theta)} \ge 0$.

Two dimensionless factors $r$ and $scal$ have been multiplied to amplify the
size of the image, otherwise which will be infinite small, since
the distance $D$ between the central compact object and
the observer and the size of the target object $L$ satisfy $D\gg L$.

In the rest frame of the observer, the spacetime is locally flat, we still have
\begin{equation}
\label{squaremomentum}-{p'}_{(t)}^2+{p'}_{(r)}^2+{p'}_{(\theta)}^2+{p'}_{(\phi)}^2=0.
\end{equation}
Using equations (\ref{alpha}), (\ref{beta}) and (\ref{squaremomentum}) and noting the signs of $p'_{(a)}, \alpha, \beta$, we obtain
\begin{eqnarray}
\label{prpt}\frac{p'_{(r)}}{p'_{(t)}}=-\frac{1}{\sqrt{1+(\alpha/r\,scal)^2+\left(\beta/r\,scal\right)^2}},\\
\frac{p'_{(\theta)}}{p'_{(t)}}=-\frac{\beta/r scal}{\sqrt{1+(\alpha/r\,scal)^2+\left(\beta/r\,scal\right)^2}},\\
\label{pppt}\frac{p'_{(\phi)}}{p'_{(t)}}=-\frac{\alpha/r
scal}{\sqrt{1+(\alpha/r\,scal)^2+\left(\beta/r\,scal\right)^2}}.
\end{eqnarray}
Substituting equations (\ref{prpt})-(\ref{pppt}) into (\ref{pivspt}),
we get the functions $\bar{p}_{(i)}(\alpha,\beta)/\bar{p}_{(t)}(\alpha,\beta)$. We can then calculate
$\lambda$ and $q$ from impact parameters by using
equations (\ref{lambda}) and (\ref{q}).

One can verify directly that when $r_{obs} \rightarrow \infty$, $scal=1$, and $\upsilon_r= \upsilon_\theta=
\upsilon_\phi=0$, the equations (\ref{lambda}) and
(\ref{q}) reduce to (\ref{alpha_inf}) and (\ref{beta_inf}) immediately.

If the observer has motion, the image on the plate will have a displacement compare to the image when
the observer is stationary. The displacement is
proportional to the observer's velocity and can be described by $\alpha_c$, $\beta_c$, which is
the coordinates of image point of the origin of B-L coordinate system on the photographic plate.
Obviously $\alpha_c$ and $\beta_c$ satisfy the following equations:
\begin{eqnarray}
\bar{p}_{(\theta)}(\alpha_c,\beta_c)=0,\\
\bar{p}_{(\phi)}(\alpha_c,\beta_c)=0.
\end{eqnarray}
Actually, $\bar{p}_{(\phi)}(\alpha, \beta)=0$ represents the projection of the spin axis of the black hole onto the
plate. When $\upsilon_r=\upsilon_\theta=\upsilon_\phi=0$, the above equations become $\alpha_c=0$ and
$\beta_c=0$. The region of the image on the plate therefore is
$[-\Delta L+\alpha_c,\Delta L+\alpha_c]$ and $[-\Delta L+\beta_c,\Delta L+\beta_c]$,
where $\Delta L$ is the half length of the image.

\subsection{Redshift formula}
The redshift $g$ of a photon is defined by $g=E_{obs}/E_{em}$. From
above discussion, we know that $E_{obs}=-p'_{(t)}$,
$E_{em}=-p_{\mu}u^{\mu}_{em}$, where $u^{\mu}_{em}$ is the
four-velocity of the emitter. If we define
\begin{equation}
f_t = \alpha_t^{(t)}+\alpha_t^{(i)}\frac{p'_{(i)}}{p'_{(t)}}.
\end{equation}
From equation (\ref{energy_t}), we have $-p'_t=Ee^{-\nu}(1-\lambda \omega)/f_t$. Using the equation (\ref{fourp}),
then $g$ can be expressed as follows
\begin{equation}
\label{redshift1}g=\frac{\left[e^{-\nu}(1-\lambda\omega)/f_t\right]_{obs}}{\left[u^t
\left(1+s_r\dot{r}\sqrt{R}/\Delta+s_{\theta}\dot{\theta}\sqrt{\Theta_\theta}-\lambda\Omega\right)\right]_{em}},
\end{equation}
where $\dot{r}=u^r/u^t,\dot{\theta}=u^\theta/u^t, \Omega=\dot{\phi}=u^\phi/u^t$ are
the coordinate velocities.

With $\dot{r}, \dot{\theta}, \Omega$, the physical velocities
of the emitter with respect to the LNRF $\upsilon_r, \upsilon_\theta, \upsilon_\phi$
can be written as \citep{bardeen1972}
\begin{equation}
\upsilon_r=e^{\mu_1-\nu}\dot{r},\quad
\upsilon_\theta=e^{\mu_2-\nu}\dot{\theta},\quad
\upsilon_\phi=e^{\psi-\nu}(\Omega-\omega),
\end{equation}
with which the four-velocity of the emitter can be expressed as
\begin{equation}
u^\mu_{em}=\gamma(e^{-\nu},\upsilon_re^{-\mu_1},\upsilon_\theta
e^{-\mu_2},\Omega e^{-\nu}).
\end{equation}
Then the $g$ can be rewritten as \citep{muller}
\begin{equation}
\label{redshift}g=\frac{\left[e^{-\nu}(1-\lambda\omega)/f_t\right]_{obs}}{\left[\gamma
e^{-\nu}\left(1+s_r e^{\nu}
\upsilon_{r}\sqrt{R}/\sqrt{\Sigma\Delta}+s_{\theta}e^{\nu}\upsilon_{\theta}\sqrt{\Theta_\theta}/\sqrt{\Sigma}-\lambda\Omega\right)\right]_{em}}.
\end{equation}
For an emitter movies in a Keplerian orbit, the formula of $g$ reduces to
\begin{equation}
\label{redshift2}g=\frac{\left[e^{-\nu}(1-\lambda\omega)/f_t\right]_{obs}}{\left[\gamma
e^{-\nu}\left(1-\lambda\Omega\right)\right]_{em}}.
\end{equation}

\section{A brief introduction to the code}
\label{genmet}
\subsection{The four coordinates and affine parameter functions}
\label{fourcoordinates}
We have expressed the four coordinates
$r$, $\mu$, $\phi$, $t$ and the affine parameters $\sigma$ as functions of $p$. We denote them as follows:
\begin{equation}
\label{rtpt}r(p),\mu(p),\phi(p),t(p),\sigma(p).
\end{equation}
In practical applications, we are interested in determining the original
position where the photon was emitted or the regions traveled by the photon.
To make the calculations effectively, all photons are traced backward from
the observer to the emitter
along the geodesics. But not all photons start from the
observer will go through the emission region one interested, and the tracing process will be terminated
either these photons go to the infinity or fall into the event horizon of a black hole.

Now we discuss how to determine the intersection of a geodesic with the surface of
an optically thick emission region, we assuming that the optical depth of which is so large that
a sharply emission surface exits. And the surface is smooth and continuous and can be
described by an algebra equation:
\begin{equation}
F(r,\theta,\phi)=F_0,\quad or \quad F(x,y,z)=F_0,
\end{equation}
where $x$, $y$, $z$ are the pseudo Cartesian coordinates and defined by
\begin{equation}\label{pesudo}
x=\sqrt{r^2+a^2}\sin\theta\cos\phi,\,\,
y=\sqrt{r^2+a^2}\sin\theta\sin\phi,\,\, z=r\cos\theta.
\end{equation}
In some special cases the surface one considered may not keep stationary, the
surface equation will be a function of time $t$, i.e., $F(r,\theta,\phi,t) = F_0 $.
We introduce a function $f(p)$ defined by
\begin{equation}
f(p)=F[r(p),\theta(p),\phi(p)]-F_0.
\end{equation}
Then the roots of equation $f(p) = 0$ correspond to the intersections of the
geodesic with the target surface. Therefore if equation $f(p)=0$ has no roots, the
geodesic will never intersect with the surface. To solve this equation effectively, we classify
geodesics into four classes denoted by A, B, C and D, according to their relationships with respect to a shell,
shown in Figure \ref{abcd}.
The shell includes the emission region completely, and its inner and outer radius
are $r_{in}$ and $r_{out}$.
Reminding that $r_{tp_1}$ and $r_{tp_2}$ are turning points, between which the radial motion
of a photon is confined. Geodesics in the four classes satisfy the conditions
A: $r_{tp_1}>r_{out}$, B: $r_{in}\le r_{tp_1}\le r_{out}$, C: $r_h< r_{tp_1}< r_{in}$ and
D: $r_{tp_1}\le r_h$ respectively, where $r_h$ is the radius of the event horizon.

The values of parameter $p$ corresponding to
the intersections of the geodesic with the shell are denoted by $p_1$, $p_2$, $p_3$, $p_4$, which satisfy
$p_1<p_2<p_3<p_4$. Obviously the roots of equation $f(p) = 0$ may exist on intervals
$[p_1,p_2]$ and $[p_3,p_4]$. We use the Bisection
or the Newton-Raphson method to search the roots \citep{numrecipes}.

Solving the radiative transfer equation in optically thin or
thick media, one needs to evaluate integrations along geodesic with taking the affine parameter
$\sigma$ as the independent variable. Since we have taken $p$ to be the independent variable,
we can replace $\sigma$ by $p$ to evaluate these integrals.
From the definition of $p$, i.e., equation (\ref{pdefine1}), one has
\begin{equation}
dp = \pm \frac{dr}{\sqrt{R(r)}}=\pm\frac{d\theta}{\sqrt{\Theta_{\theta}}},
\end{equation}
and from the equations (\ref{defr}) and (\ref{deftheta}) one gets
\begin{equation}
d\sigma = \pm\Sigma \frac{dr}{\sqrt{R(r)}}=\pm \Sigma \frac{d\theta}{\sqrt{\Theta_\theta}}.
\end{equation}
From above equations one immediately obtains
\begin{equation}
\label{aff_p}d\sigma= \Sigma dp,
\end{equation}
which converts the independent variable from $\sigma$ to $p$ in radiative transfer applications \citep{Yuan2009}.

Finally we give a brief discussion on the determination of a
geodesic connecting the emitter and observer
\citep{viergutz1993, beckwithdone2005}.
We use $\alpha_{em}$, $\beta_{em}$ to represent the
impact parameters of the geodesic which connecting the observer and emitter and $p_{em}$ to indicate the position of the emitter
on the geodesic, in which the coordinates of the emitter are $r_{em}$, $\mu_{em}$ and $\phi_{em}$. Obviously we have the
following set of equations:
\begin{eqnarray}
\label{obsemit1}r(p_{em},\alpha_{em},\beta_{em})&=&r_{em},\\
\label{obsemit2}\mu(p_{em},\alpha_{em},\beta_{em} )&=&\mu_{em},\\
\label{obsemit3}\phi(p_{em},\alpha_{em},\beta_{em} )&=&\phi_{em}.
\end{eqnarray}
In principle if we can solve this set of nonlinear equations simultaneously for $p_{em},\alpha_{em},\beta_{em}$,
the geodesic is determined uniquely. Therefore the observer-emitter problem also becomes a root finding
problem. In our code we use the Newton-Raphson
method \citep{numrecipes} to solve these equations.
\subsection{The code}
\label{code}
In this section we shall give a brief introduction for the
code, and a more detailed introduction is given in the README\footnote{\url{http://www1.ynao.ac.cn/~yangxl/readme.pdf}}
file. The code is named YNOGK (Yun-Nan Observatory Geodesics Kerr)
and written by Fortran 95, in which
the object-oriented method has been used. The code is composed by
a couple of modules. For
each module, a special function has been implemented and one can use all
supporting functions and subroutines in that module by a command "use
module-name" in his/her own program. By adding corresponding
modules into one's own code, one can easily develop new ones to handle special and
more sophisticated applications.

Two modules named {\bf{ell-function}} and {\bf{BLcoordinate}} are the most important ones in
ynogk, the former one includes supporting functions and subroutines for calculating the
Carlson's elliptic integrals and the R-functions. Many routines in this
module come from geokerr \citep{dexagol2009} and Numerical recipes \citep{numrecipes}.
The latter module includes routines for computing all
coordinates and the affine parameter functions: $t(p),r(p),\theta(p),\phi(p)$ and $\sigma(p)$.
To call these routines, constants of motion and components of four-momentum $\bar{p}_{(r)}$,
$\bar{p}_{(\theta)}$, $\bar{p}_{(\phi)}$ measured in a LNRF reference
must be prescribed, which can be computed by two subroutines named {\bf{lambdaq}} and {\bf{initialdirection}}
in ynogk. The former routine computes the constants of motion from impact parameters, while the latter one
computes $\lambda$ and $q$ from the initial $p'_{(a)}$ given in a reference $K'$, which has physical
velocities $\upsilon_r, \upsilon_\theta, \upsilon_\phi$ with respect to the LNRF.
Of course one can compute them by his/her own subroutines according to their needs.
According to the discussion in Section \ref{fourcoordinates}, we present a module named
{\bf{pem-finding}} to search the minimum root $p_{em}$ of equation $f(p) = 0$, and $f(p)$
as an external function should be given by the user.
In module {\bf{obs-emitter}}, we present routines to find the root of equations (\ref{obsemit1})-(\ref{obsemit3})
by using the Newton-Raphson algorithm \citep{numrecipes}. In the testing section of the code, this module has
been used to determine the geodesic connecting the observer and the central point of a
hot spot, which movies in the inner most stable circular orbit (ISCO). With the geodesic the motion of the spot
can be described easily. The results are agree very
well with previous works, in which a very different method has been used to determine the
motion of the spot, i.e., by tabulating the motion according to
the time of the observer over one period \citep{schnittman2004,dexagol2009}.

All routines for computing the Carlson's elliptic integrals have been extensively checked
by NIntegrate function of Mathematica. The original code of these routines comes
from geokerr \citep{dexagol2009}, and has been modified to adapt to our code.
The original code for the computing of R-functions comes from Numerical recipes
\citep{numrecipes}. The same
check also has been done for the functions $t(p),r(p),\theta(p),\phi(p), \sigma(p)$. When
$|\alpha|$ and $|\beta|$  $\lesssim10^{-7}$, we let them to be zero, since the Carlson's integrals can not
maintain their accuracy. The treatment is same
for any other parameters if they take offending values. For some critical cases, special
treatments also have been implemented.

\subsection{Comparisons and speed tests}
Our code has many common points with geokerr of \cite{dexagol2009}. We both use the Carlson's
method to compute the elliptic integrals and use the elliptic functions
to express all coordinates as functions of a independent variable.
But the elliptic functions we used are mainly the Weierstrass' elliptic function $\wp(z;g_2,g_3)$,
which has a cubic polynomial, leading simpler root distribution and the cases of integral are reduced.
In our code the four B-L coordinates $r$, $\theta$, $\phi$, $t$ and
affine parameter $\sigma$ are expressed as analytical or numerical functions of a
parameter $p$, which corresponds to $I_u$ or $I_\mu$ in \cite{dexagol2009}.
With this treatment one can compute the geodesics directly without providing any information about
the turning points in advance. Which also allows one to
track emissions from a more sophisticated surface, not only for standard thin accretion disk.
In the code testing section we will show the images of a warped disk,
which has a curved surface.

Our strategy, i.e., expressing coordinates as functions of $p$ semi-analytically, can be extended to
compute the timelike geodesics directly, almost without any modifications.
As mentioned in \cite{dexagol2009},
the calculations of the timelike geodesics involve many more cases.
The main challenge is to specify the number of radial turning points for bounded orbits
in advance. But our strategy does not require the specification of the number of turning points both in
radial and poloidal coordinates in advance, therefore which can be used naturally and effectively in the calculations of
timelike geodesics even in a Kerr-Newmann spacetime.

In our code we give the orthenormal tetrad of the emitter or the observer analytically, provided
the physical velocities of which with respect to the LNRF are specified. They may be useful
in Monte-Carlo type code of radiative transfer, which needs one to make transformations from
the reference of the emitter to the B-L coordinate system frequently \citep{Dolence2009}.
As illustrated in figure \ref{equatorialplane}, emissions in the reference of the emitter
are specified isotropically, but from the perspective of the B-L coordinate system which are anisotropic
due to the Doppler beaming effect.

\begin{figure}[ht!]
\begin{center}
\includegraphics[width=0.6\textwidth]{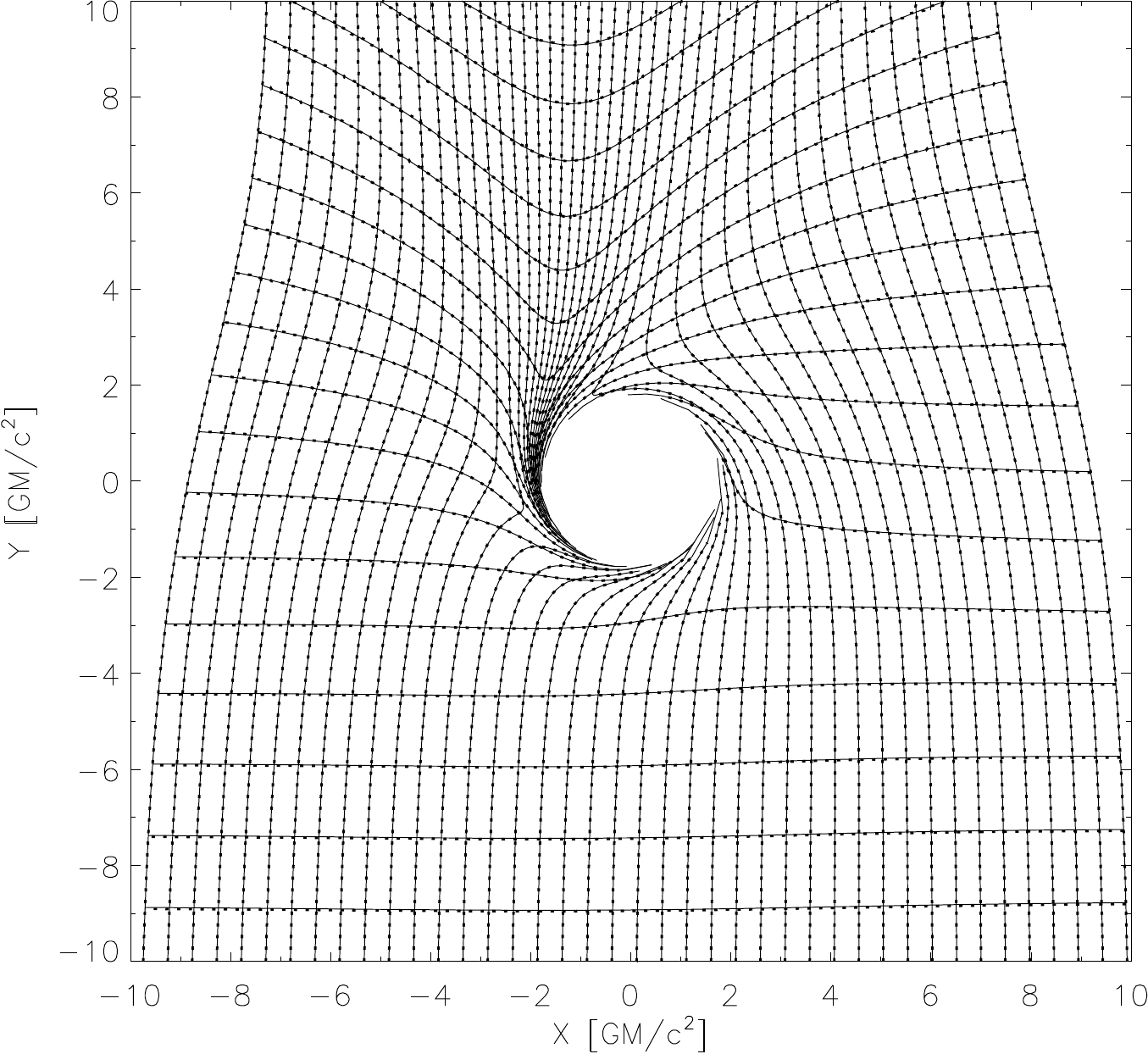}
\caption{\label{projections} The projection of
a uniform grid from the photographic plate of the observer onto the equatorial plane of a black hole is shown. The inclination
angle $\theta_{\mathrm{obs}}$ is $60^\circ$ and the black hole spin a is 0.95. Solid lines represent the results from
our code and the dotted lines from geokerr.  x and y
are pseudo-Cartesian coordinates in the equatorial plane of the black hole.}
\end{center}
\end{figure}

\begin{figure}[ht!]
\begin{center}
\includegraphics[width=0.5\textwidth]{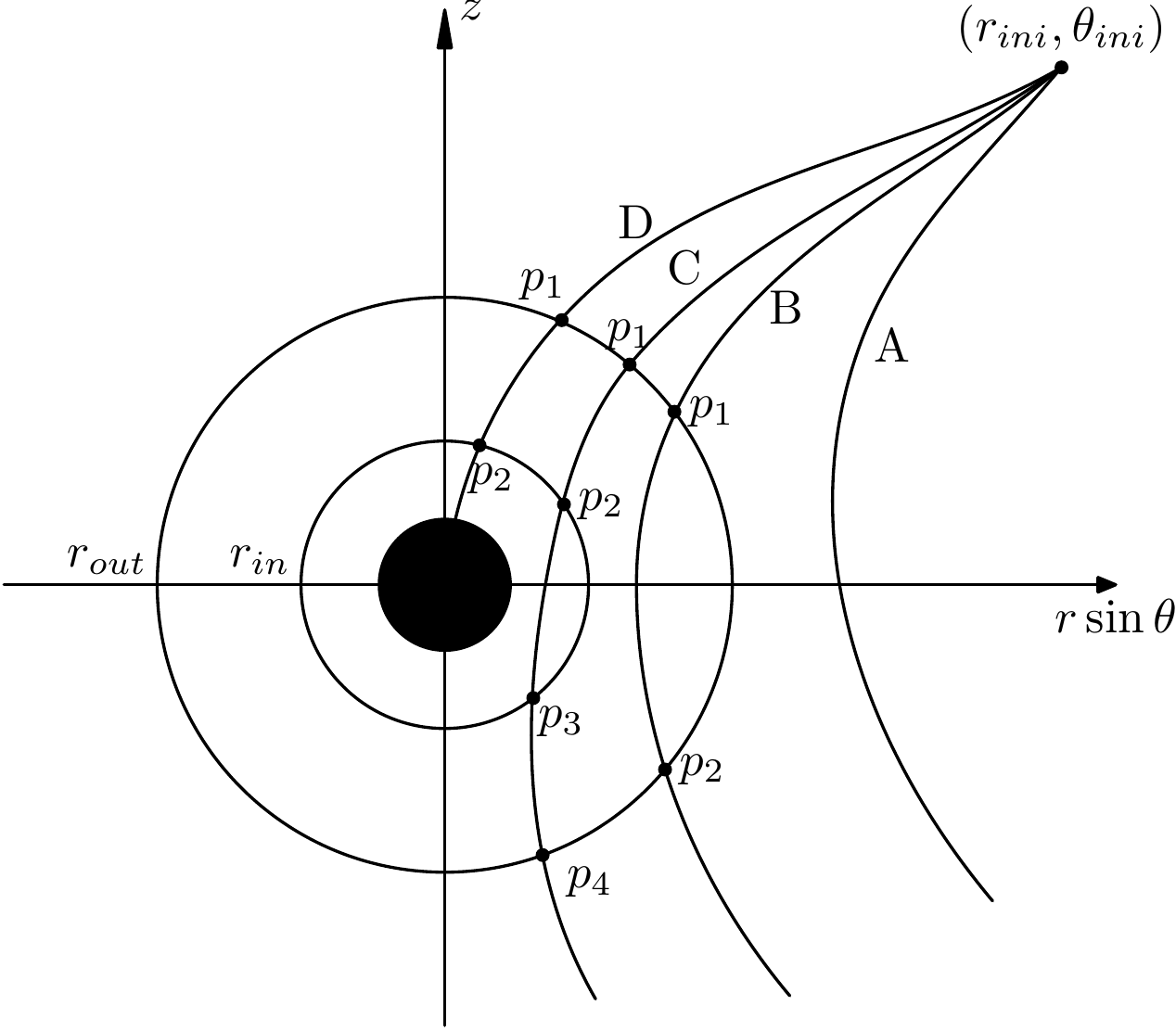}
\caption{\label{abcd}
The classifications of a set of null geodesics according to their relationships with respect to
a shell, the inner and outer radius of which are $r_{in}$ and $r_{out}$ respectively. The geodesics are classified
into four classes, marked by A, B, C and D. Since the target object or the emission region are assumed to be completely
included by the shell, only geodesics in classes B, C and D
have probabilities to intersect with the target object or go through emission region.
The trajectories of the geodesics are schematically plotted and have been projected onto the $r$-$\theta$ plane.
The central black region represents the black hole shadow.}
\end{center}
\end{figure}

Our testing results for various toy problems agree well with those of \cite{dexagol2009}.
In Figure \ref{projections} we illustrate the projection of a uniform orthonormal grid from the photographic plate of the
observer onto the equatorial plane of a black hole, in which the solid and dotted lines represent the results from
our code and geokerr respectively. They agree with each other very well.

The basic strategy used in ynogk to compute the elliptic integrals and functions are very similar to
geokerr. For example we make ynogk to compute the minimum number of R-functions possible
and share them between routines. This strategy improves the speed of our code greatly.
But there is still some differences between the two codes. Firstly, we assemble $r(p)$ and
$\mu(p)$ into routines for computing $t(p)$, $\phi(p)$ and $\sigma(p)$, thus the repeated calculations
for the same integrals among those functions can be avoided. We
provide a routine named {\bf{ynogk}} to
compute the four B-L coordinates and affine parameter simultaneously.
We also provide two independent routines named {\bf{radius}} and {\bf{mucos}} to compute
$r(p)$ and $\mu(p)$ respectively. Secondly, ynogk can save the values of variables
used in the calculations for a same geodesic but for different $p$. These values
can be used repeatedly.

ynogk has almost same speed with the geokerr in tracing radiations from an
optically and geometrical thin disk, because there is only
one point, i.e., the intersection of the ray with the disk surface, needs to be calculated.
For the calculations of radiative transfer, in which many points are needed along each geodesics,
ynogk has a little slower than geokerr. The speed tests of a code are not only dependent
on the applications mostly, but also on the environment of the code running.
From the testing results of ynogk,
we expect that the speed of which is almost same with geokerr in many other applications.

For more detailed introductions, one can see the README file. In
the next section, we will show the testing results of our code for toy problems.

\section{The tests of our code}
\label{protest} In this section, we shall present the testing results of our code
for toy problems. The results not only demonstrate the validation our code,
but also give specific examples of its utility.
Firstly, we show the image of a black hole
shadow, in which the intensity represents the value of the affine parameter
$\sigma$. By this example we want to test the validation of function $\sigma(p)$.
Then we show the images of a couple of accretion disks
and a rotationally supported torus, all of them are optically thick and have a sharply emission
surface. The disks include the standard thin, thick and
warped disks. Next we show the images of a ball orbiting around
a Kerr black hole in a Keplerian orbit to illustrate the gravitational lensing effect. Then we calculate
the line profiles of the Fe K$\alpha$ and the blackbody radiation spectra of a standard thin accretion
disk around a Kerr black hole. In order to test the accuracy of function $t(p)$, we
image a hot spot moving around a Kerr black hole in the ISCO
for various black hole spins. We also calculate the spectrogram and light curves of
the spot over one period of the motion with various inclinations for a Schwarzschild black hole. Finally we discuss the
radiative transfer equation and its solution, whit which the radiative transfer
process in a radiation dominated torus around a black hole has been discussed. We give the
images of the torus for optically thin and thick cases. The resolution of images in this section is
taken to be $801\times801$, each pixel corresponds to an unique geodesic.

\subsection{Black hole shadow}
\begin{figure}[ht!]
\begin{center}
\includegraphics[width=0.8\textwidth]{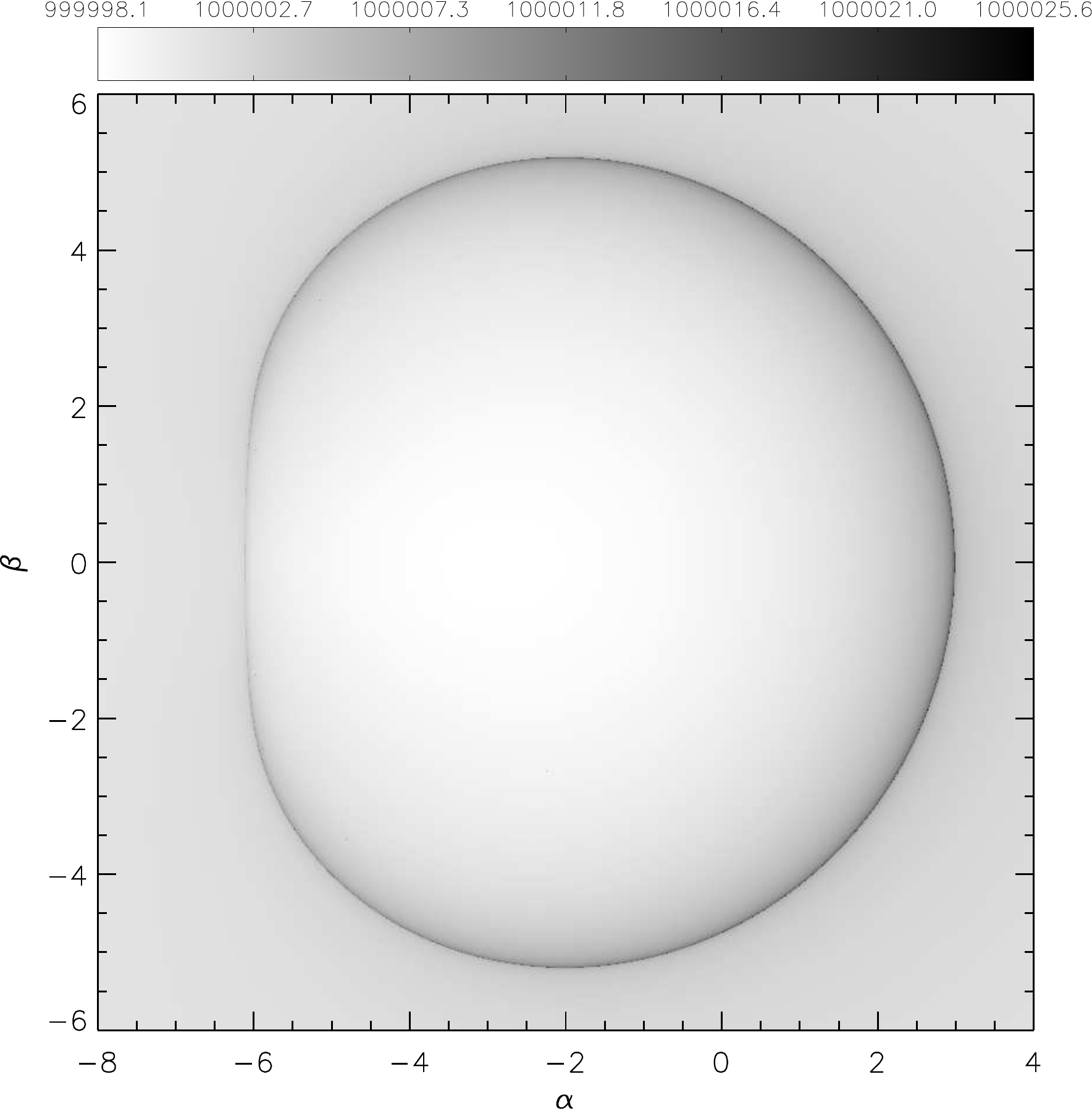}
\caption{\label{shadow}The shadow
of a black hole with near extremal spin ($a=0.998$) from the edge-on view is shown. The
radial coordinate of the observer is $10^6$ $r_g$.
The greyscale represents the value of the affine parameter $\sigma$ evaluated from the
observer to the terminated position---either at the black hole or re-emerging to
the starting radius.
Compare to Figure 2 of
\citet{dexagol2009}.
$\alpha$ and $\beta$ are the impact parameters, which describe the size and
the position of the image on the photographic plate.}
\end{center}
\end{figure}

As the first test of our code we give the image of a black hole shadow. We trace all
photons backward from the photographic plate to the black hole along geodesics.
The intensities of the image are taken to be the affine parameter $\sigma$
evaluated from the observer to terminations on the geodesic---either when it intersects
with the event horizon of the black hole or reaches a turning point and returns to
the starting radius. The evaluations of the affine parameter outside the shadow are multiplied a
factor 1/2. In Figure \ref{shadow}, we show the image from an edge-on view.  We take the spin $a$ to
be 0.998, and the distance of the observer to be $10^6$ $r_g$, where
$r_g$ is the gravitational radius.

To evaluate affine parameter $\sigma$ from function $\sigma(p)$, we need $p_h$, which is the value
of parameter $p$ corresponds to the event horizon and also the root of equation $r(p)=r_h$.
We can get $p_h$ by evaluating the integral of $r$ in the definition of $p$, and need
not to solve this equation directly. We provide
a routine named {\bf{r2p}} to complete this evaluation.

\subsection{Accretion disks}
Next we present the images of the accretion disks around a Kerr
black hole, including the standard thin, thick and warped disks.
The imaging of the disks is usually taken as the first step to
calculate the line profiles of the Fe K$\alpha$ and the spectrum
\citep{li2005}. Usually the pseudo colors of the image represent the redshift $g$ or the
observed flux intensity $I_{\nu}$ of emissions come from the disk.

\begin{figure}[ht!]
\begin{center}
\includegraphics[width=1.\textwidth]{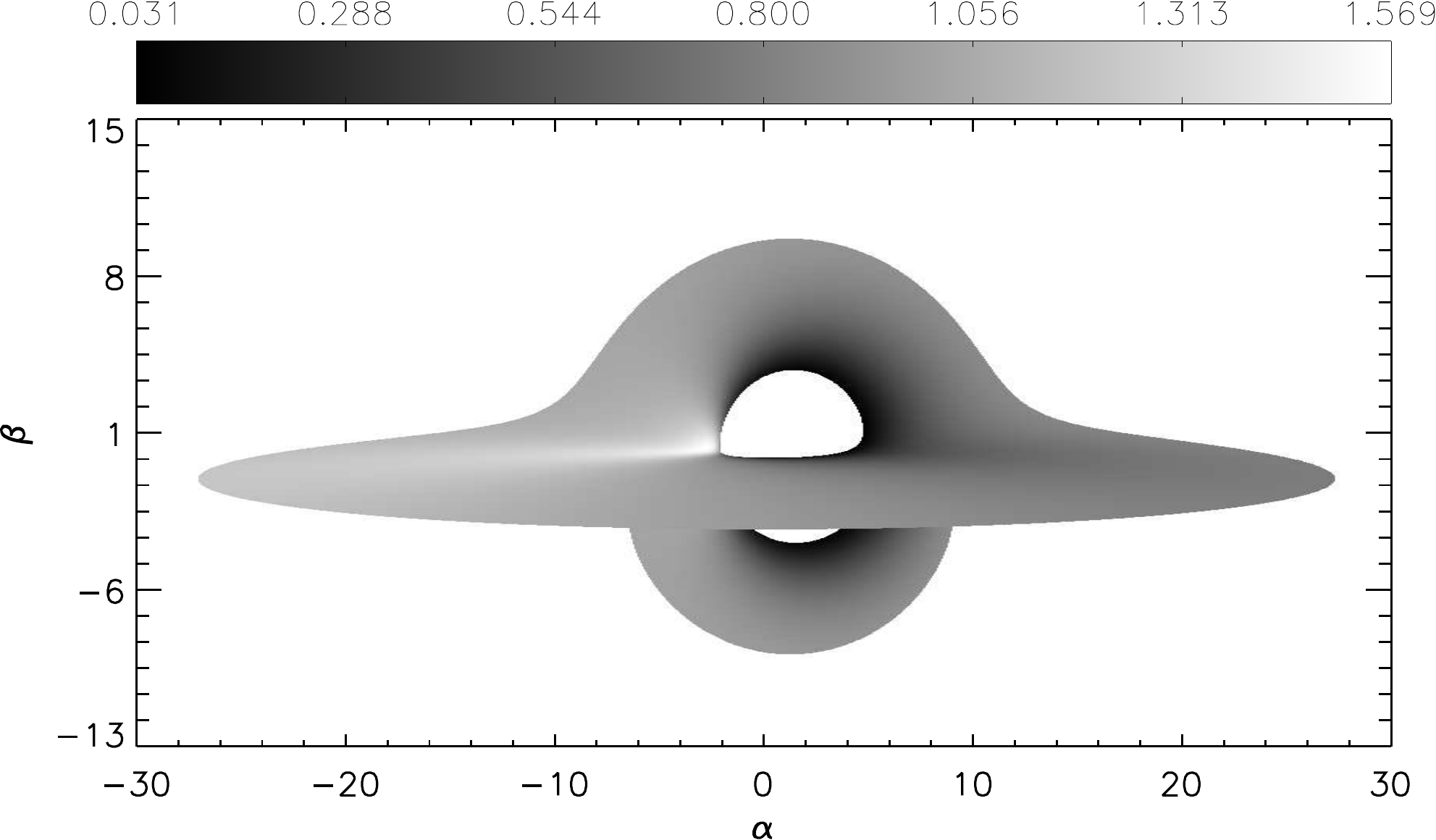}
\caption{\label{thindisk}
This figure shows the image of a
standard thin accretion disk, whose inner and outer radius are $r_{\mathrm{ms}}$ and 22
$r_g$ respectively. The black hole spin $a$ is 0.998 and
the inclination angle $\theta_{obs}$ is $86^\circ$. The
radial coordinate of the observer is 40 $r_g$. One can see Figure 6 of
\citet{beckwithdone2005} or Figure 4 of \citet{dexagol2009} for
comparison. The high-order image is also shown.
$\alpha$ and $\beta$ are the impact parameters, and
the intensity of the greyscale represents the redshift $g$ of emissions come
from the surface of the disk.}
\end{center}
\end{figure}

In Figure \ref{thindisk}, we show the image of a standard thin disk, the inner and
outer radius of which is $r_{ms}$ and 22 $r_g$ repectively. The black
hole spin $a$ is 0.998 and the inclination angle $\theta_{obs}$
is $86^\circ$. The distance of the observer is
40 $r_g$. The shape of the image is quite different from
the one observed from infinite far away.
We also illustrate the high-order images of the disk in
this figure. Due to the light bending and focusing, one can see the part
behind the black hole and the bottom side of the disk.
The color intensities represent the redshift $g$ of
emissions come from the disk.

In order determine the intersections of geodesics with the disk,
we need to know the minimum root of equation $\mu(p) = 0$. We provide two routines named
{\bf{pemdisk}} and {\bf{pemdisk-all}} to compute the root by evaluating the integral of $\mu$ in
the definition of $p$. Using
{\bf{pemdisk}} one can draw the direct image, while using {\bf{pemdisk-all}} one can
draw the direct and high-order images.

\begin{figure}[ht!]
\begin{center}
\includegraphics[width=1.0\textwidth]{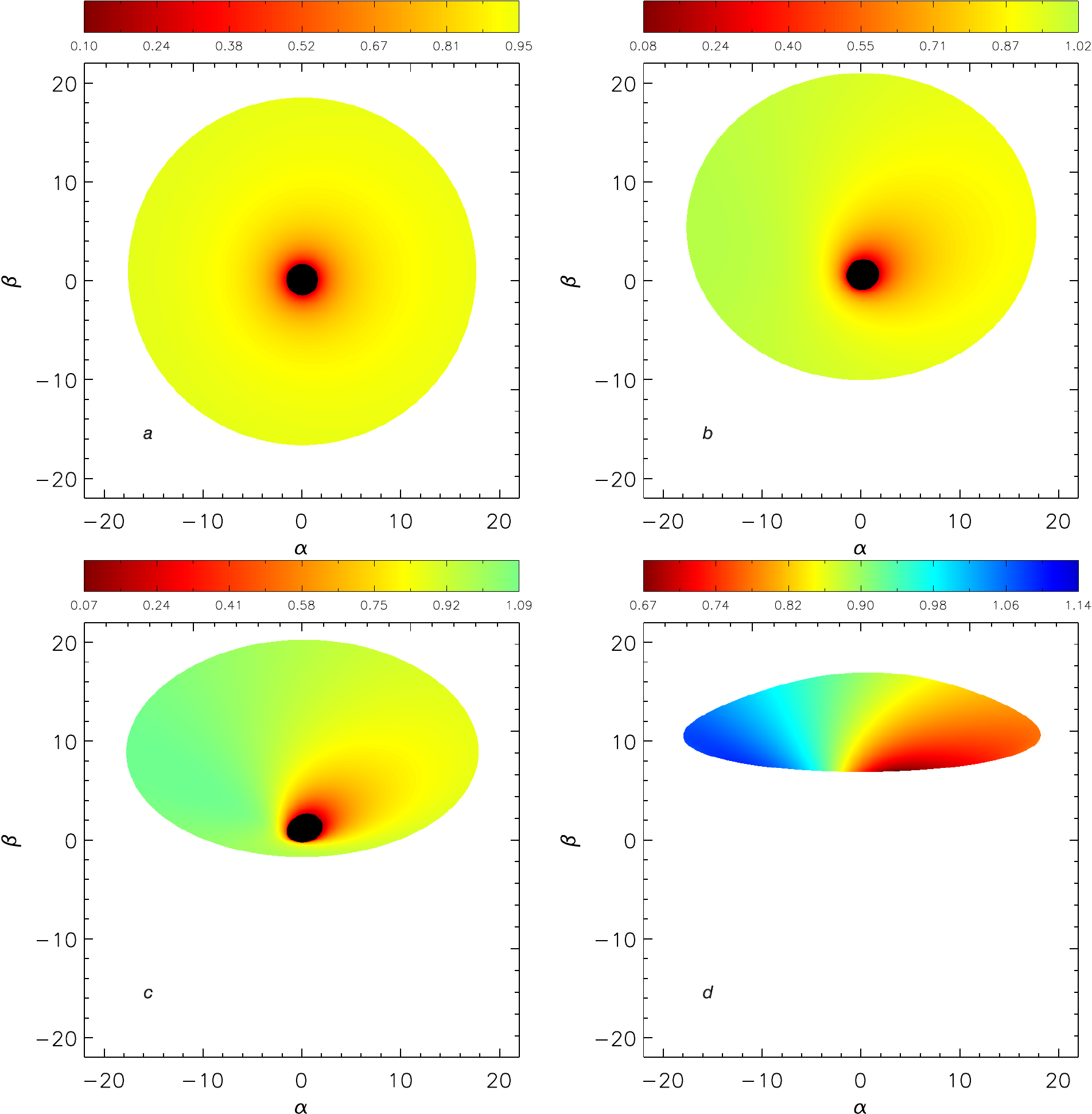}
\caption{\label{thickdisk}
Images of a thick disk around a near extremal Kerr black hole (a=0.998) for various
inclination angles are shown.
The surface of the disk has a constant inclination angle $\delta$ with respect to
the equatorial plane and $\delta$ is taken to be $30^\circ$. The inner radius is
the marginally stable circular orbit $r_{\mathrm{ms}}$ and the outer radius is 20 $r_g$. The inclination angles
$\theta_{obs}$ are $5^\circ$, $30^\circ$, $55^\circ$ and $80^\circ$
for panels a, b, c and d respectively. $\alpha$ and $\beta$ are the impact parameters, and
the intensities of the color represent the redshift $g$ of emissions come
from the surface of the disk. Compare to Figure 10 of
\citet{wu2007}.}
\end{center}
\end{figure}

The surface of the thick disk has a constant inclination angle $\delta$ with respect to
the equatorial plane \citep{wu2007}. To trace the thick disk, we need to
solve equation $\mu(p)=\cos(\pi/2-\delta)$ to get $p$ for the upper surface and
$\mu(p)=\cos(\pi/2+\delta)$ for the bottom surface, the roots of these two equations can
also be computed by {\bf{pemdisk}} and {\bf{pemdisk-all}}. Since the surface
particles of the disk no longer keep in the equatorial
plane, they will do the sub-Keplerian motion with a angular velocity
given by \citep{ruszkowskifab2000}
\begin{equation}
\Omega=\left(\frac{\theta}{\pi/2}\right)^{\frac{1}{n}}\Omega_K+
\left[1-\left(\frac{\theta}{\pi/2}\right)^{\frac{1}{n}}\right]\omega,
\end{equation}
where $\Omega_K=1/(r^{3/2}+a)$ is the Keplerian velocity and parameter $n$ is
taken to be 3 here. Using
the equation (\ref{redshift2}), we can calculate the redshift of the emissions come from the
disk. The images are shown in Figure \ref{thickdisk},
which agrees very well with Figure 10 of \cite{wu2007}.

\begin{figure}[ht!]
\begin{center}
\includegraphics[width=0.8\textwidth]{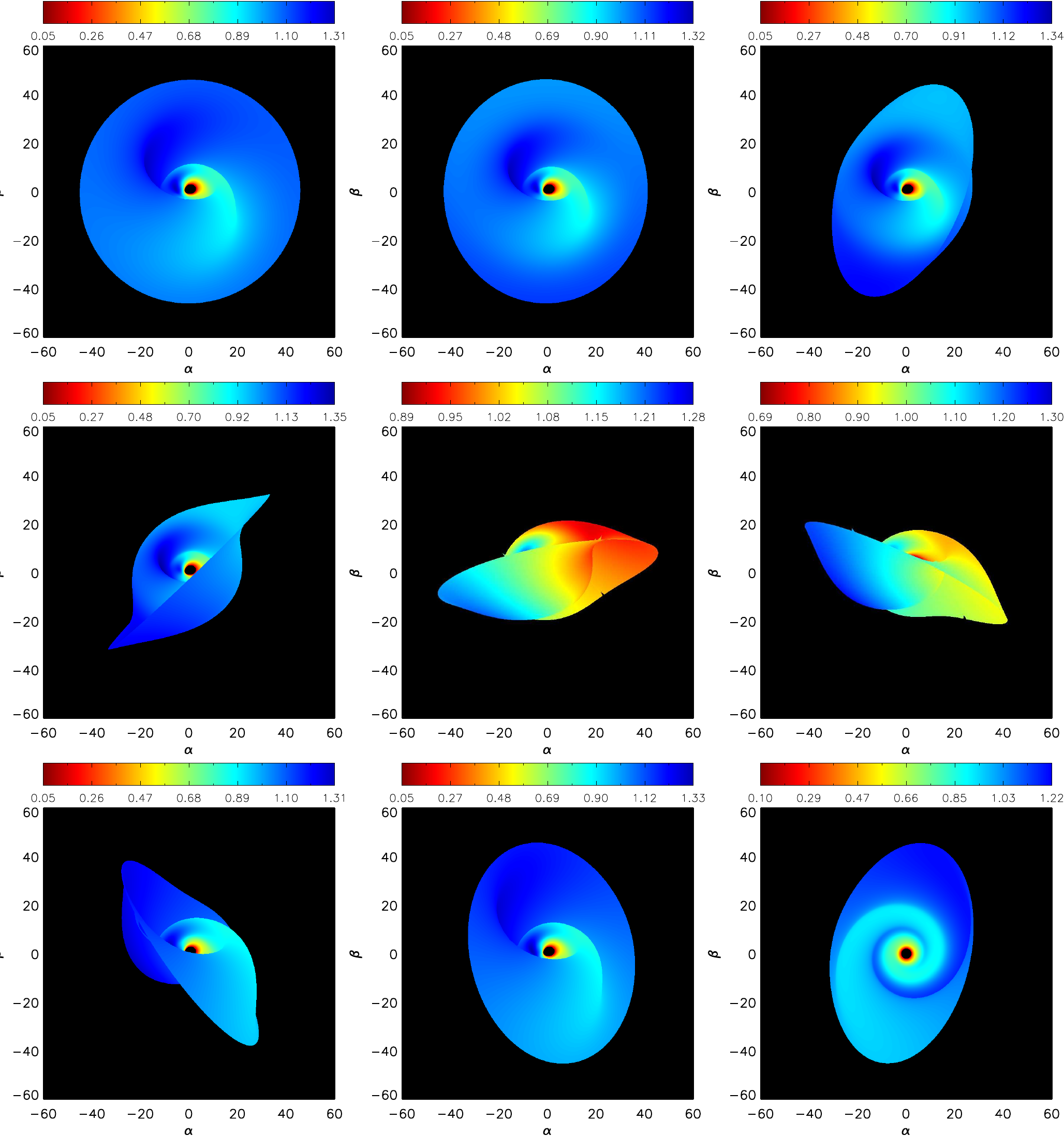}
\caption{ \label{warpeddisk}
This figure shows the images of a warped accretion disk around a near extremal black hole (a=0.998) viewed from
different azimuthal angles.
The inner and outer radius of the disk are $r_{ms}$ and 50 $r_g$. The observer's inclination angle $\theta_{obs}$ is 50$^\circ$.
The warping parameters are $n_1=4\pi$, $n_2=4$,
and $n_3=0.95$. The azimuthal angle $\gamma_0$, which represents the view angle, is $0^\circ$,
$45^\circ$, $90^\circ$, $135^\circ$, $180^\circ$, $225^\circ$, $270^\circ$ and $315^\circ$
for panels from left to right and top to bottom respectively. For comparison we show a image observed from a
face-on view in the final panel. The false color also represents
the redshift $g$ of the emissions come
from the surface of the disk. $\alpha$ and $\beta$ are the impact parameters.
We take the parameter $n_1\neq 0$, leading the warping of the disk
along the azimuthal direction shown clearly in the final panel, which is the main difference
compare to Figure 3 of \cite{wang2012}.}
\end{center}
\end{figure}

The warped accretion disk is also a very interesting object in astrophysics
\citep{bardeen1975,wu2007,wang2012}. Here we discuss a very simple model for the warped disk, in which
the disk is assumed to be optically thick and its surface can be described by \citep{wang2012}
\begin{equation}
\cot\theta=-\tan\beta \cos(\phi-\gamma),
\end{equation}
where parameters $\gamma$ and $\beta$ are defined by
\begin{eqnarray}
\gamma(p)&=&\gamma_0+n_1\exp\left[n_2\frac{r_{in}-r(p)}{r_{out}-r_{in}}\right],\\
\beta(p)&=&n_3\sin\left[\frac{\pi}{2}\frac{r(p)-r_{in}}{r_{out}-r_{in}}\right],
\end{eqnarray}
where $r_{in}$ and $r_{out}$ are the inner and outer radius of the disk, and
$n_1$, $n_2$, $n_3$ are the warping parameters.
With above equations, we get the $f(p)$ as follows
\begin{equation}
f(p) = \tan\beta(p) \cos[\phi(p)-\gamma(p)]+\frac{\mu(p)}{\sqrt{1-\mu^2(p)}}.
\end{equation}
With the minimum root of equation $f(p)=0$, we can image the warped disk. For
the poloidal velocity $\dot{\theta}$ of the particle is
nonzero, the formula (\ref{redshift1}) or
(\ref{redshift}) is used to calculate the redshift g (cf. \citet{wang2012}). The images of the
warped disk are shown in Figure \ref{warpeddisk}, in which the warping parameters $n_1$ and $n_2$ are nonzero,
leading the disk warps along azimuthal direction. For comparison one can see Figure 3 of \citet{wang2012}, in
which $n_1$ is taken to be zero for simplicity, thus the shape of the disk is quite different from the one illustrated
here.

\subsection{Rotationally supported torus}
\label{torus_image}
\begin{figure}[ht!]
\begin{center}
\includegraphics[width=0.8\textwidth]{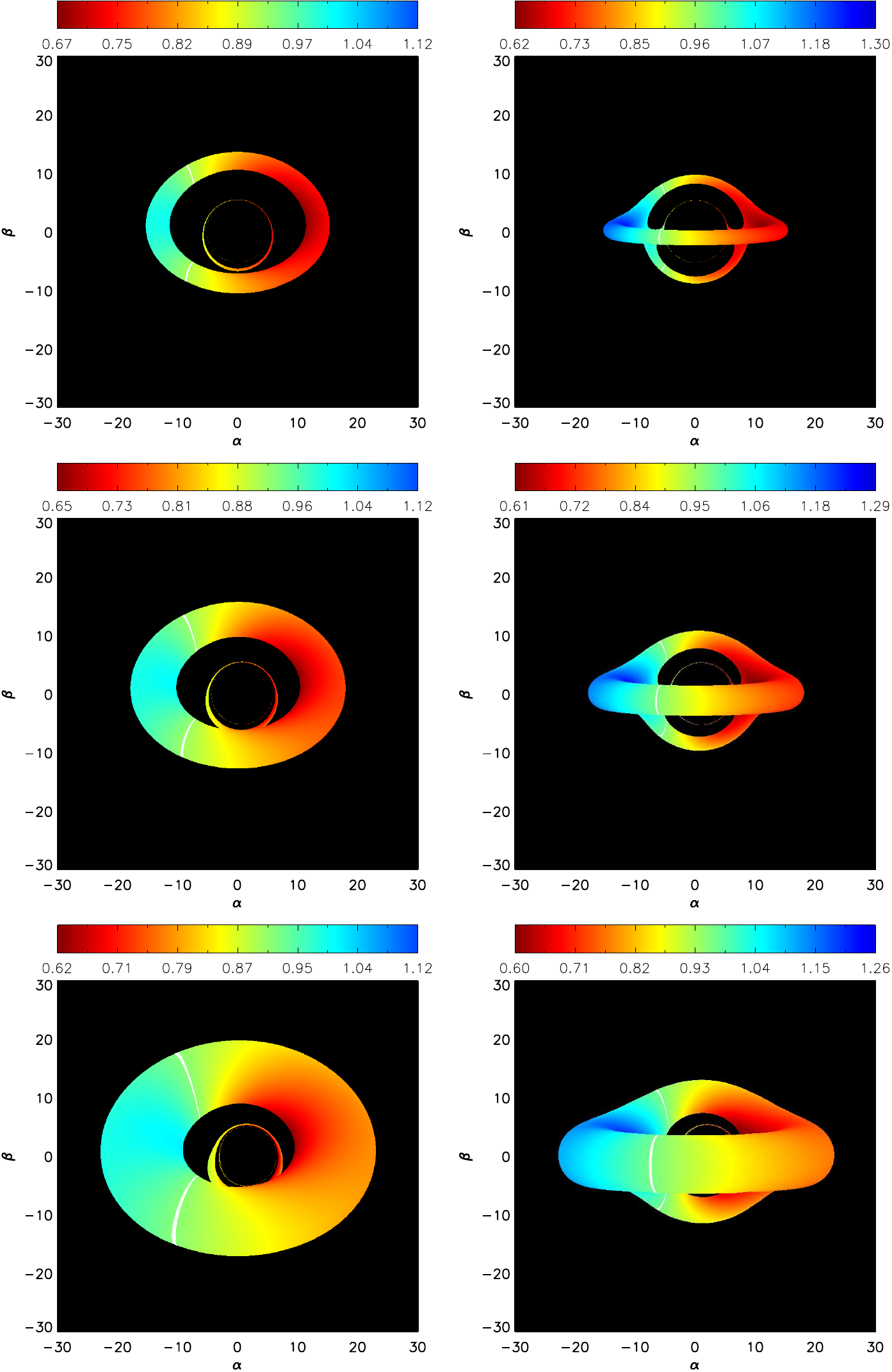}
\caption{\label{torus}
Images of a rotationally supported torus, which is geometrical and optically thick, are shown.
The torus parameters are $n=0.2$, $r_k = 12$ $r_g$. The black hole spin $a$ is
0, 0.5 and 0.998 for panels from top to bottom. The
inclination angle $\theta_{obs}$ is $45^\circ$ for left column and $85^\circ$ for
right column. The false color represents
the redshift $g$ of the emissions come
from the surface of the torus and the white areas represent the
zero-shift regions. $\alpha$ and $\beta$ are the impact parameters.
Compare to Figure 3 of \cite{Younsi2012}.}
\end{center}
\end{figure}

In this section, we give the images of a rotationally supported torus.
For simplicity we give a brief introduction for the torus model here,
for the more detailed discussions one is recommended to the paper of \cite{fuerstwu2004} or
\cite{Younsi2012}. The torus is assumed to be stationary and axisymmetric. Due to the balance of the
centrifugal force, gravity and pressure force, the structure of the torus
is stratified and the isobaric surfaces can be described by a set of
differential equations \citep{Younsi2012}
\begin{eqnarray}
\label{torus_isobar1}\frac{dr}{d\zeta} = \frac{\psi_2}{\sqrt{\psi_2^2+\Delta \psi_1^2}},\\
\label{torus_isobar2}\frac{d\theta}{d\zeta} = \frac{-\psi_1}{\sqrt{\psi_2^2+\Delta \psi_1^2}},
\end{eqnarray}
where
\begin{eqnarray}
\label{torus_contour1}\psi_1 & =& M\left(\frac{\Sigma-2r^2}{\Sigma^2}\right)\left(\Omega^{-1}-a\sin\theta\right)^2+r\sin^2\theta,\\
\label{torus_contour2}\psi_2 &=& \sin2\theta\left(\frac{Mr}{\Sigma^2}[a\Omega^{-1}-(r^2+a^2)]^2+\frac{\Delta}{2}\right),\\
\label{torus_omega}\Omega &=& \frac{\sqrt{M}}{(r\sin\theta)^{3/2}+a\sqrt{M}}\left(\frac{r_k}{r\sin\theta}\right)^{n},
\end{eqnarray}
and $\Omega$ is the angular velocity. $r_k$ represents the radius at which the
the particle orbits with a Keplerian velocity. The index
parameter $n$ is crucial for regulating the angular velocity profile and
adjusting the geometrical aspect ratio of the torus. $\zeta$ is an auxiliary
parameter. In order to give the
outer surface of the torus, one needs to specify the most inner radius of the
torus, which is usually regarded as the intersection of the isobaric surface
with the ISCO. Taking the inner most radius in the equatorial plane to
be the initial condition, the differential equations (\ref{torus_isobar1}) and
(\ref{torus_isobar2}) are now readily to be integrated. In Figure \ref{torus}
we illustrate the images of the torus, which has the same
parameters with Figure 3 of \cite{Younsi2012}, the results agree with
each other very well.

\subsection{The gravitational lensing effect}
\begin{figure}[ht!]
\begin{center}
\includegraphics[width=1.0\textwidth]{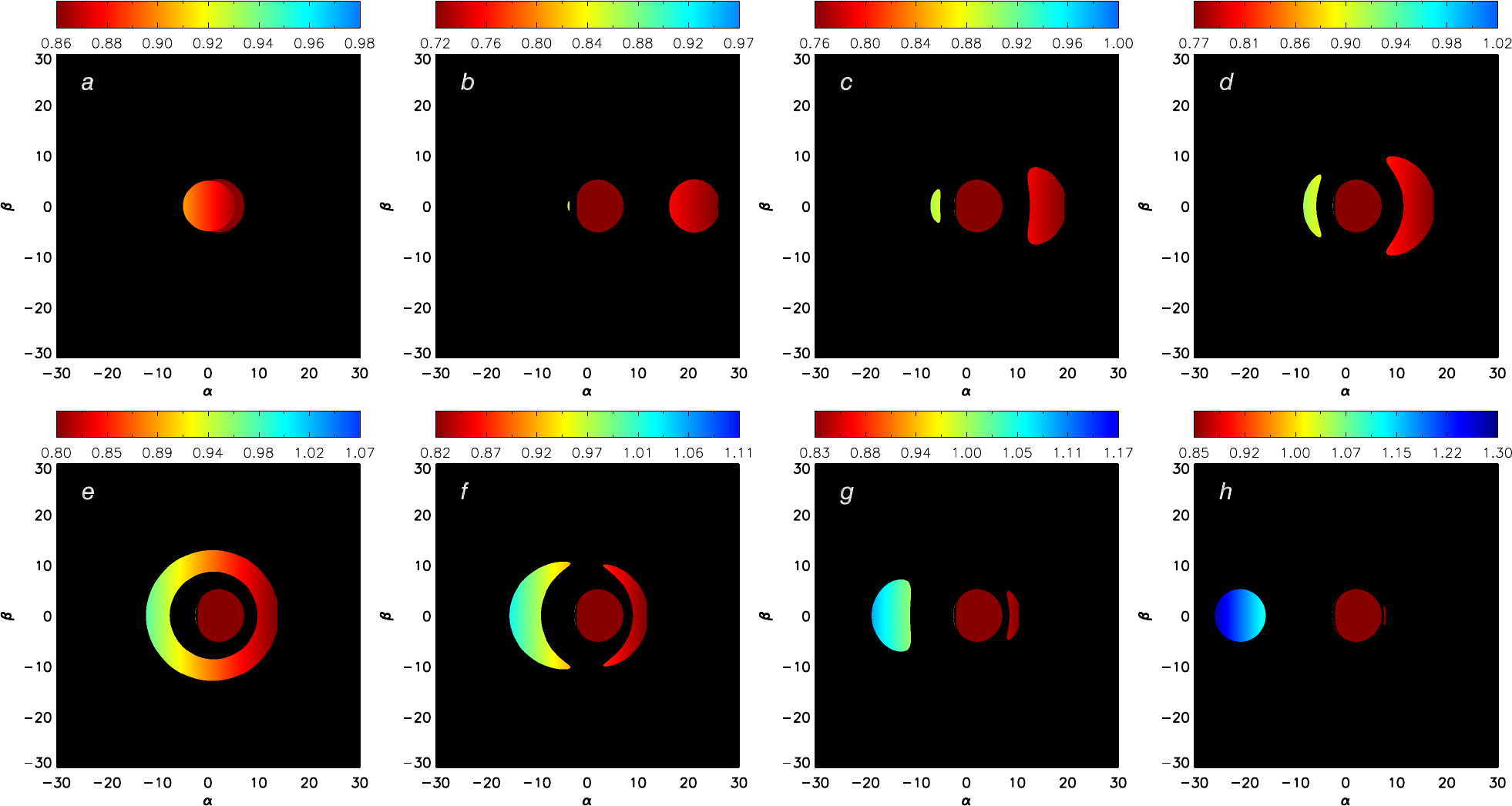}
\caption{\label{weaklense}
This figure illustrates the gravitational lensing effect by the motion of a ball
movies around a near extremal black hole (a=0.998) in a Keplerian orbit.
The motion is observed from an edge-on view. The radius of the ball and the orbit are
5 $r_g$ and 20 $r_g$ respectively. The central red
region represents the black hole shadow. The azimuthal angles of the ball
measured from the line of sight along inverse-clockwise direction are
$0^\circ,90^\circ,150^\circ,160^\circ,180^\circ,195^\circ,210^\circ,270^\circ$
for panels a-h. The pseudo
color shows the redshift of the emissions come from the surface of the
ball. $\alpha$ and $\beta$ are the impact parameters.}
\end{center}
\end{figure}

Due to the strong gravity field, when the trajectory of a photon is closed to
the vicinity of a compact object, it will be bent or focused, then multiple images
will be observed, this is the so called gravitational lensing effect. Here this phenomenon will
be illustrated by a simple example, in which a ball
moves around a near extremal black hole (a=0.998) in a Keplerian orbit. The radius of the
orbit is $R_0$, then the angular velocity of the ball is $\Omega = 1/(R_0^{3/2}+a)$.
The coordinates of the center of the ball will be
\begin{eqnarray}
&&x_0(p) = \sqrt{R_0^2+a^2}\cos[\Omega t(p)],\\
&&y_0(p) = \sqrt{R_0^2+a^2}\sin[\Omega t(p)],\\
&&z_0(p) = 0.
\end{eqnarray}
Then the function of the surface of the ball
can be expressed as follows:
\begin{equation}
f(p)=\sqrt{[x(p)-x_0(p)]^2+[y(p)-y_0(p)]^2+z(p)^2}-r_1,
\end{equation}
where $r_1$ is the radius of the ball. The images observed from an
edge-on view are illustrated in Figure \ref{weaklense}.
For different positions of the ball in its orbit, the image changes greatly,
which even becomes a ring as the ball movies to the back of the even horizon.

\subsection{The line profiles of Fe K$\alpha$}
\begin{figure}[ht!]
\begin{center}
\includegraphics[width=0.7\textwidth]{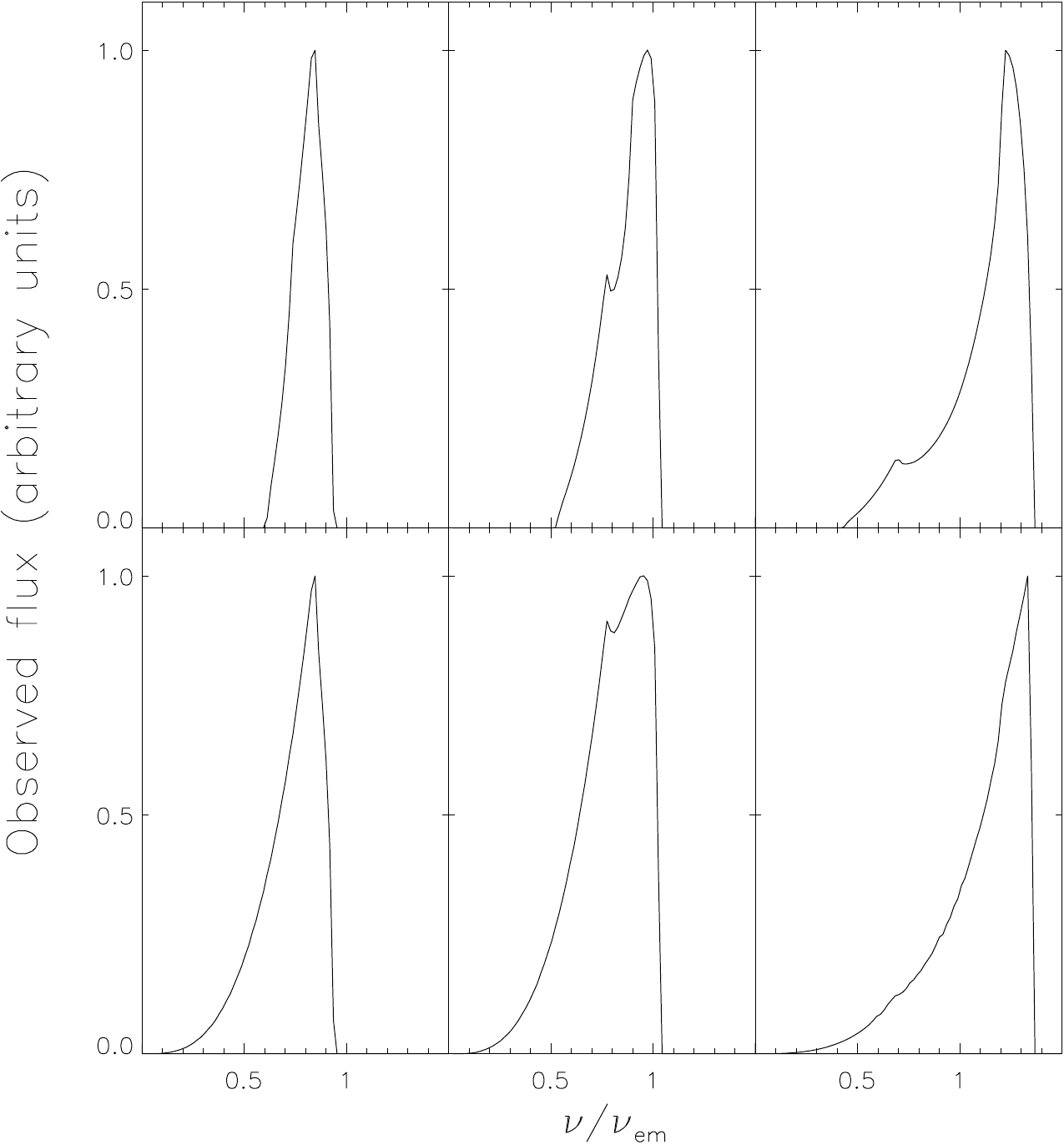}
\caption{\label{kelines}The
theoretical line profiles of the Fe K$\alpha$ of a thin accretion disk for
various black hole spins and inclination angles are shown. The inner and outer radius of
the disk are $r_{ms}$ and 15 $r_g$. Top row: $a=0.2$; bottom row; $a=0.998$. Left column:
$\theta_{obs}=10^\circ$; middle column: $\theta_{obs}=30^\circ$;
right column: $\theta_{obs}=75^\circ$. The horizonal and vertical axes represent the frequency and
flux of the line respectively and are normalized. The line profiles are in agreement well
with the Figure 3 of \citet{cadez1998}.}
\end{center}
\end{figure}

The calculation of line profiles is very easy provided the structure of the disk is specified.
For simplicity, we assume that the particles of the accretion flow do the Keplerian motion and the disk
is geometrical thin and optically thick. The inner and outer radius of the disk are
located at $r_{ms}$ and 15 $r_g$ respectively. The emission is monochromatic and the profile of which
can be described by the Dirac's $\delta$ function in the local rest frame of the flow
\begin{equation}
I_{em}(\nu)=\frac{\epsilon_0}{4\pi r^n}\delta(\nu-\nu_{em}),
\end{equation}
where $n$ is the index of emissivity and assumed to be 3. Since
$I_\nu/\nu^3$ is an invariance along a geodesic \citep{mtw}, we get the observed
intensity $I_{obs}=g^3I_{em}$, where
$g=\nu_{obs}/\nu_{em}$ is the redshift. Then the observed flux density $F_{\nu}$ at frequency $\nu$
can be computed by integrating $I_{obs}$ over the whole plate as following expression
\begin{equation}
F_{\nu}=\int\frac{\epsilon_0}{4\pi r^n}\delta(\nu-\nu_{em})g^3 d\alpha d\beta.
\end{equation}
The observed intensities have been normalized in the computation.
The results are shown in Figure \ref{kelines}, which agrees very well with the Figure
3 of \citet{cadez1998}. From this figure one can see that the higher black hole spin leads
the broadening in low frequency for the ISCO is closer to the event horizon, the gravitational
redshift effect is remarkable, while the higher inclination angle leads the
broadening in high frequency for the Doppler beaming effect.

\subsection{The blackbody radiation spectrum of a Keplerian disk}
In this section we will compute the spectrum of a Keplerian disk around a Kerr black hole
to illustrate effects of the black hole spin and the observer's inclination
angles on the observed profiles of the spectrum \citep{li2005}.
Similarly, the disk is assumed to be geometrical thin and optically thick, and the radiation spectrum of the disk
in its local rest frame is an isotropic blackbody spectrum. We denote the effective temperature of the disk by $T_{\mathrm{eff}}$.
Then radiation intensity at frequency $\nu$ can be written as
\begin{equation}
I_{em}(\nu)=\frac{h\nu^3}{\exp(h\nu/k_{\mathrm{B}}T_{\mathrm{eff}})-1},
\end{equation}
where $h$ and $k_{\mathrm{B}}$ are the Plank and Boltzmann constants respectively. For a blackbody radiation,
the effective temperature is simply
\begin{eqnarray}
T_{\mathrm{eff}}=\left[\frac{F(r)}{\sigma_{SB}}\right]^{1/4},
\end{eqnarray}
where $\sigma_{SB}$ is the Stefan-Boltzmann constant. Here we do not consider effect of
the returning radiation of the disk on the spectrum, therefore $F(r)$ is just the energy flux emitted from the disk's surface
measured by a locally corotating observer.
For the Keplerian accretion disk around a Kerr black hole, \cite{page1974} have get the analytical expression
for $F(r)$, i.e.,
\begin{eqnarray}
F(r)=\frac{\dot{M}}{4\pi r}f,
\end{eqnarray}
where $\dot{M}$ is mass accretion rate, $f$ is a function of $r,a,r_{ms}$, and the seminal expression
of which is given by equations (15d) and (15n) of \cite{page1974}.
With $F(r)$ the effective temperature of the disk can be computed readily.
Using the invariance $I_\nu/\nu^3$, one can get the observed intensity $I_{obs}=g^3I_{em}$.
The total observed flux density at frequency $\nu$ therefore is the integration of
$I_{obs}$ over the whole plate
\begin{equation}
F(\nu_{obs})=\int \frac{h\nu_{obs}^3 }{\exp(h\nu_{obs}/gk_\mathrm{B}T_{\mathrm{eff}})-1}d\alpha d\beta.
\end{equation}
Then the photon number flux density is $N_{obs}=F(\nu_{obs})/E_{obs}$.

\begin{figure}[ht!]
\begin{center}
\includegraphics[width=0.6\textwidth]{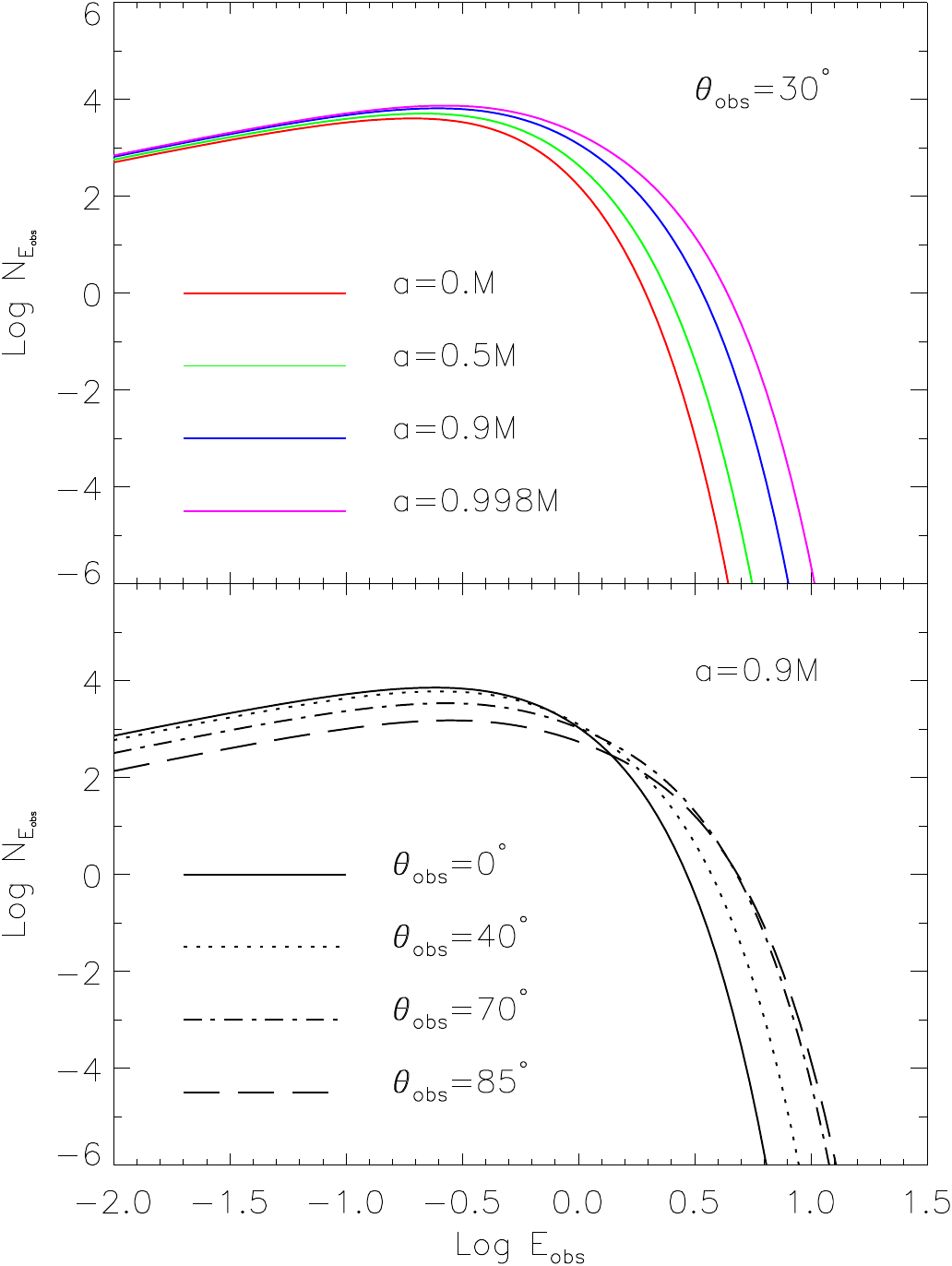}
\caption{\label{spectrum}The
effects of the black hole spin (top) and the inclination angle (bottom) on
the spectrum of a standard thin accretion disk around a Kerr black hole are
shown. The inner and outer radius of
the disk are $r_{ms}$ and $30$ $r_g$, where $r_{ms}$ is the radius of marginally stable
circular orbit. }
\end{center}
\end{figure}

The results are plotted in Figure \ref{spectrum}. Compare to the Figure 5 of \citet{li2005} we find that the basic
features of the two figures are in agreement. For example, as shown in top panel, we see that as the spin of the black hole goes up the
spectrum becomes harder. Physically, this is due to the fact that as the spin $a$ increases, the system of the
accretion disk has a higher radiation efficiency and a higher temperature. In the bottom panel of Figure \ref{spectrum}, we
can see that at the low-energy end, the flux density goes down as $\theta_{obs}$ goes up. As explained by
\cite{li2005} this is caused by the projection effect. While at the high-energy end, the flux density goes up as the
$\theta_{obs}$ increases. As pointed out by \cite{li2005} this is resulted from the joint action of
the effects of Doppler beaming and gravitational focusing.

In the top panel there is a noticeable effect: even though we do not consider the returning radiation, the flux density
goes up as the spin increases in the low-energy end. \cite{li2005} suggested that this effect is caused by the
returning radiation. We proposal that this explanation may be not correct and the effect is just caused by the
simple fact that a higher spin leads to a higher radiation efficiency and temperature.

\subsection{The motion of a hot spot}
In order to test the validation of the function $t(p)$ and illustrate the time
delay effect in the Kerr spacetime, we image a hot
spot orbiting around a black hole retrogradely in a marginally stable circular orbit
for various spins and compute the observed light curve and spectra. The radius of
the spot is $R_{\mathrm{spot}}$=0.5 $r_g$. The emissivity of the spot is taken to
be Gaussian shape in its rest frame \citep{schnittman2004}, i.e.,
\begin{equation}
j(\mathbf{x})\propto
\exp\left[-\frac{|\mathbf{x}-\mathbf{x}_{\mathrm{spot}}(t)|^2}{2
R_{\mathrm{spot}}}\right].
\end{equation}
For the motion of the spot, one must consider the time
delay effect and the azimuthal position of the spot when imaging the
spot and calculating its spectra.
In order to compute the time delay $\Delta t$ for each geodesic starting from the photographic
plate, a reference time $t_{\mathrm{obs}}$, which is taken to be the time used by a
photon traveling from the central point of the spot to the observer, needs to be specified.
Meanwhile the position of the spot can be determined by its central coordinates
($r_{ms}$, $\mu=0$, $\phi_{\mathrm{spot}}$).
Then with the method discussed in section \ref{genmet}, (i.e., the method to
determine a geodesic connecting the observer and emitter with the given coordinates), we
can determine the geodesic connecting the central point of the spot and the observer.
With this geodesic the reference time $t_{\mathrm{obs}}$ can be calculated readily.
Using $t_{\mathrm{obs}}$, we can easily calculate the time delay $\Delta t$ for each geodesic, and
$\Delta t= t_{\mathrm{geo}}-t_{\mathrm{obs}}$, where $t_{\mathrm{geo}}$ is the time used by a photon traveling from
the observer to the disk following the geodesic.
With $\Delta t$ and the position of the spot, we can
compute the distance between the intersection of the geodesic with the disk
and the center of the spot, i.e., $|\mathbf{x}-\mathbf{x_{\mathrm{spot}}}|$.
Thus the emissivity can be computed readily.

\begin{figure}[ht!]
\begin{center}
\includegraphics[width=0.8\textwidth]{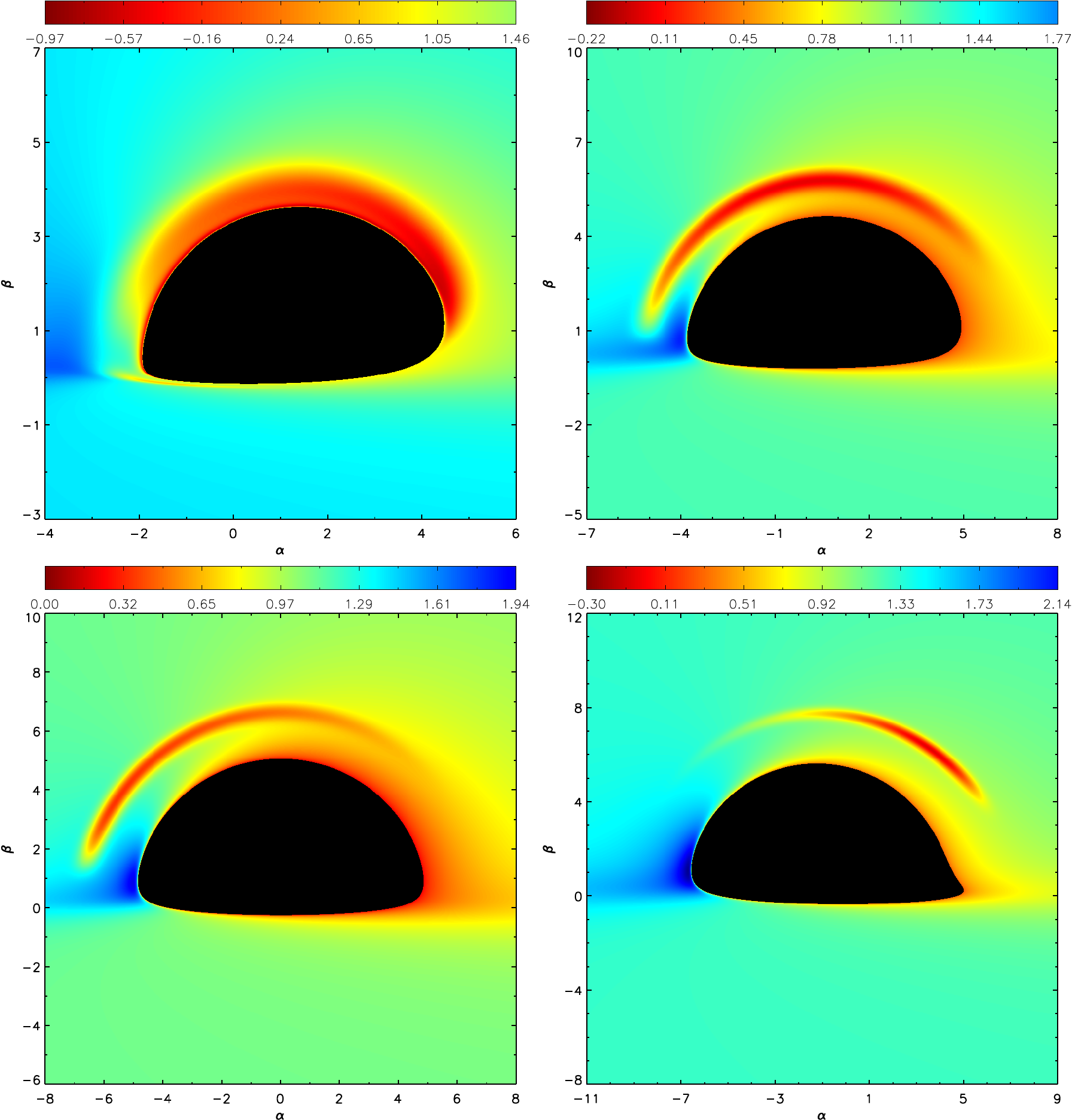}
\caption{\label{spot}
The images of a hot spot orbits around a black hole for different black hole spins are shown.
The spot lies in a standard thin accretion disk and its central point
is fixed at the ISCO. We have extended the
inner radius of the disk to the photon orbit $r_{ph}$, at which the energy per
unit rest mass of a particle is infinite. It is also
the innermost boundary of circular orbit for particles
\citep{bardeen1972}. For the panels from left to right and top to bottom, the black hole spin a is
0.998, 0.5, 0 and -0.998 respectively. The inclination angle
$\theta_{obs}$ is $85^\circ$. The false color represents the value of $g-j(\mathbf{x})$,
where $g$ is the redshift of the emissions come from the surface of the disk, and
$j(\mathbf{x})$ is the emissivity of the spot.}
\end{center}
\end{figure}

In Figure \ref{spot}, we illustrate the images of the spot with different black hole spins.
As the spin increases, the marginally stable circular orbit is
closer to the event horizon of the black hole, and the time delay effect becomes
remarkable. The image of the spot is seriously warped,
especially when the spot movies to the back of the event horizon.

\begin{figure}[ht!]
\begin{center}
\includegraphics[width=0.6\textwidth]{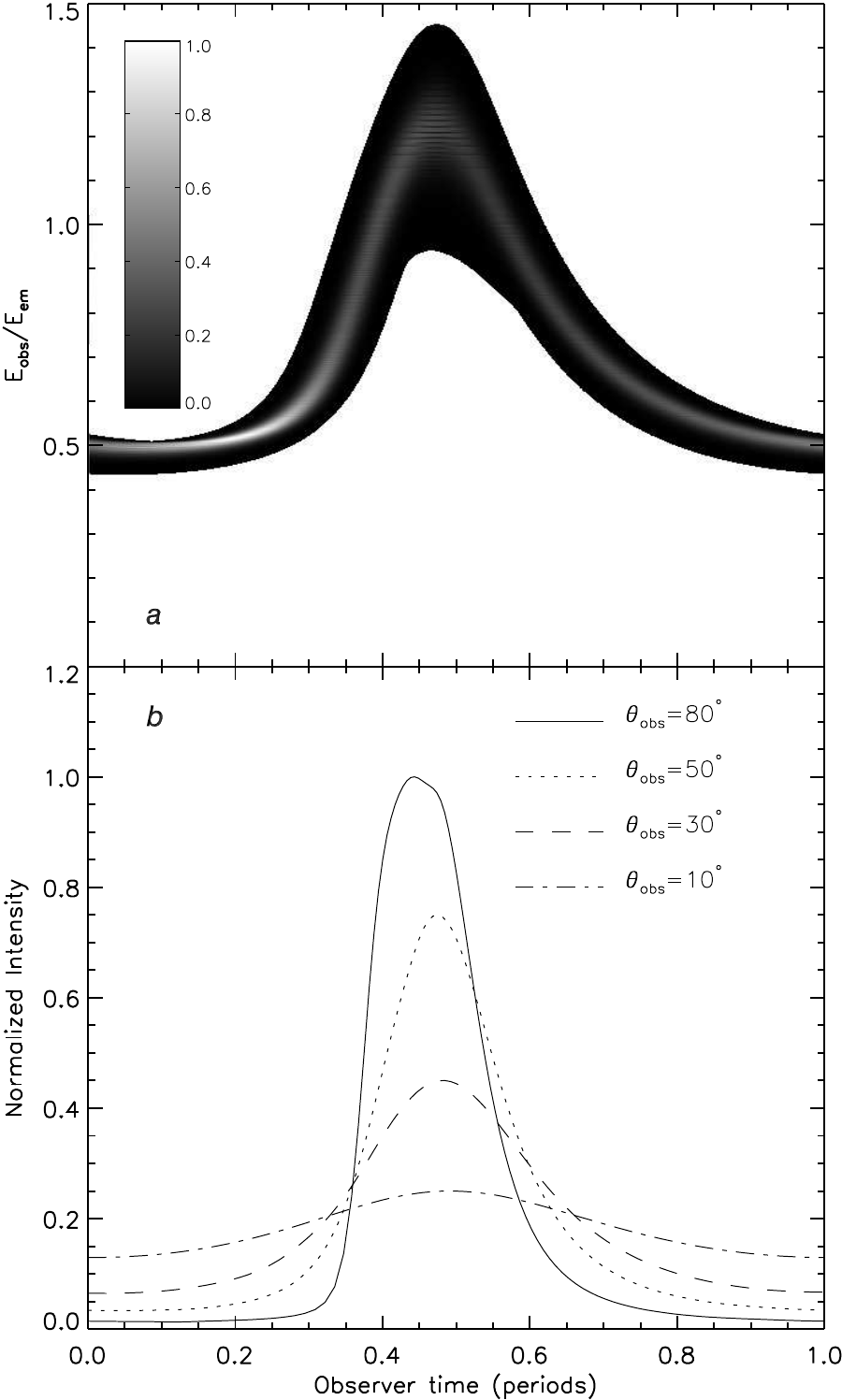}
\caption{\label{spectrum_spot}
The time-dependent spectrogram (panel a) and light curves (panel b)
of a hot spot orbits around a Schwarzschild black hole in the
marginally stable circular orbit (6 $r_g$) over one period are shown.
The inclination angle $\theta_{\mathrm{obs}}$ is $60^\circ$ for the spectrum.
The greyscale in panel a represents total sum of emissivity
$j(\mathbf{x})$ of emissions which are observed at the same time and have the same redshift $g$.
The greyscale has been normalized
and the maximum is taken to be 1.
Compare to Figure 6 and 7 of \citet{dexagol2009}.}
\end{center}
\end{figure}

When an image is obtained, the redshift and Gaussian
emissivity of all points on the spot can be computed.
Repeating this procedure over one period of the motion gives a time-dependent
spectrum. Integrating the spectrum over frequency, or equivalently over
the impact parameters, gives the light curve. The spectrum and light curve are
shown in Figure \ref{spectrum_spot}, which agree well with the results shown in Figures 6 and 7 of
\citet{dexagol2009}.

\subsection{Radiative transfer}
\subsubsection{The radiative transfer formulation}
In this section we give a brief discussion to the radiative transfer process under the Kerr spacetime.
One can find more detailed discussions
from \cite{fuerstwu2004} and \cite{Younsi2012}. It is well known
that $\mathcal{I} = I_\nu/\nu^3$, $\chi=\nu \alpha_\nu$ and $\eta=j_{\nu}/\nu^2$
are Lorentz invariants, where $I_\nu$ is the specific intensity of the radiation, $\alpha_\nu$ and
$j_{\nu}$ are the absorption and emission coefficients at the frequency $\nu$.
The radiative transfer equation reads \citep{Younsi2012}
\begin{equation}
\frac{d\mathcal{I}}{d\tau_\nu} = -\mathcal{I}+\frac{\eta}{\chi},
\end{equation}
where $\tau_\nu$ is the optical depth at the frequency $\nu$,
and defined by $d\tau_\nu = \alpha_\nu ds$ and $ds = -p_\mu u^\mu d\sigma$, in which
$ds$ is the differential distance element of a photon traveling in the
rest frame of the medium, $\sigma$ is the affine parameter, $p_\mu$ is the four momentum
of the photon, and $u^\mu$ is the four velocity of the medium. Then the radiative transfer
equation can be rewritten as \citep{Younsi2012}
\begin{equation}
\label{radequ}\frac{d\mathcal{I}}{d\sigma} = -p_\mu u^\mu|_\sigma\left(-\alpha_\nu \mathcal{I}+\frac{j_\nu}{\nu^3}\right).
\end{equation}
The solution of above equation is \citep{Younsi2012}
\begin{equation}
\mathcal{I}(\sigma) = \mathcal{I}(\sigma_0)e^{-\tau_\nu(\sigma)}-\int^\sigma_{\sigma_0}\frac{j_\nu(\sigma'')}{\nu^3}
\exp\left(-\int^\sigma_{\sigma''}{\alpha_\nu (\sigma') |p_\mu u^\mu|_{\sigma'}|d\sigma'}\right)
p_\mu u^\mu|_{\sigma''}d\sigma'',
\end{equation}
where the optical depth is
\begin{equation}
\tau_\nu(\sigma) = -\int^\sigma_{\sigma_0}\alpha_\nu(\sigma') p_\mu u^\mu|_{\sigma'} d\sigma'.
\end{equation}
As discussed in section \ref{fourcoordinates}, we can
convert the independent variable from affine parameter $\sigma$ to parameter $p$.
Using $\sigma=\sigma(p)$ and $d\sigma=\Sigma dp$, we can rewrite the solution as the integration of
parameter $p$ \citep{Yuan2009}
\begin{equation}
\mathcal{I}(p) = \mathcal{I}(p_0)e^{-\tau_\nu(p)}-\int^p_{p_0}\frac{j_\nu(p'')}{\nu^3}
\exp\left(-\int^p_{p''}{\alpha_\nu (p') |p_\mu u^\mu|_{p'}|\Sigma' dp'}\right)
p_\mu u^\mu|_{p''}\Sigma'' dp'',
\end{equation}
where
\begin{equation}
\tau_\nu(p) = -\int^p_{p_0}\alpha_\nu(p') p_\mu u^\mu|_{p'}\Sigma' dp'.
\end{equation}
With above formulae one can deal with radiative transfer problems without considering the
scattering contributions to the absorption and emission coefficients as did by
\cite{Yuan2009} and \cite{Younsi2012}.

\subsubsection{Radiative transfer in pressure supported torus}
\label{torus_image2}
In section \ref{torus_image} we have discussed a rotationally supported torus
and demonstrated its images. When the torus is optically thick,
only the emissions come from the boundary surface are considered.
When the torus is optically thin, all parts of the torus will do
contributions to the observed emissions.
We need to consider the radiative transfer procedure along the ray inside the torus.
To get the absorption and emission coefficients, we need to konw the structure model of the tours,
which determines the distributions of the temperature, mass density, pressure etc.

Firstly we construct the model of the torus, in which the torus is a perfect fluid and
its energy-momentum tensor is given by \citep{Younsi2012}
\begin{equation}
T^{\alpha\beta}=(\rho+P+\epsilon)u^\alpha u^\beta+P g^{\alpha\beta},
\end{equation}
where $\rho$ is the mass density, $P$ is the pressure, and $\epsilon$ is the internal
energy, $u^{\alpha}$ is the four velocity of the fluid, and $g^{\alpha\beta}$ are the
contravariant components of the Kerr metric. From the conservation law, namely
$T^{\alpha\beta}_{\quad;\beta} = 0$,
we get the equation of motion of the fluid as follows \citep{Abram1978}:
\begin{equation}
\label{torus_state_eq}\frac{\partial_\alpha P}{\rho+P+\epsilon}=-u_{\alpha{;\beta}}u^\beta,
\end{equation}
where the semicolon ; represents the covariant derivative, and
$u_{\alpha{;\beta}}u^\beta=a_\alpha$ is the four acceleration of the fluid.
For the torus is stationary and axisymmetric, we have $a_t = 0$,
$a_\phi=0$, and $a_r$, $a_\theta$ are given by \citep{Younsi2012}
\begin{eqnarray}
\label{ar}a_r & =& -\dot{t}^2\left[M\left(\frac{\Sigma-2r^2}{\Sigma^2}\right)\left(1-a\sin\theta\Omega\right)^2+r\sin^2\theta\Omega^2\right],\\
\label{at}a_\theta &=& -\dot{t}^2\sin2\theta\left(\frac{Mr}{\Sigma^2}[a-(r^2+a^2)\Omega]^2+\frac{\Delta\Omega^2}{2}\right),
\end{eqnarray}
where $\dot{t}=u^t$ is the time component of the four-velocity,
$\Omega$ is the angular velocity and takes the form of equation (\ref{torus_omega}).
They satisfy following equation
\begin{eqnarray}
u^t=\frac{1}{\sqrt{-(g_{tt}+2g_{t\phi}\Omega+g_{\phi\phi}\Omega^2)}}.
\end{eqnarray}
Since the torus is assumed to be radiation dominated,
the pressure $P$ can be
regarded as the sum of gas pressure $P_{\mathrm{gas}}$ and radiation pressure $P_{\mathrm{rad}}$, and
\begin{eqnarray}
P_{\mathrm{gas}} &=& \frac{\rho k_\mathrm{B} T}{\mu m_\mathrm{H}}=\beta P,\\
P_{\mathrm{rad}} &=& \frac{\sigma T^4}{3} = (1-\beta)P,
\end{eqnarray}
where $k_\mathrm{B}$ is the Boltzmann constant, $\mu$ is the mean molecular weight, $m_\mathrm{H}$ is
the mass of a hydrogen, $\beta$ is the ratio of gas pressure to the total pressure,
and $\sigma=\pi^2k^4/15\hbar^3c^3$ is the black-body emission constant. From the above equations,
one finally obtains
\begin{eqnarray}
\label{p_rho}P = \hbar c\left[\frac{45(1-\beta)}{\pi^2(\mu m_\mathrm{H}\beta)^4}\right]^{1/3}\rho^{4/3},\\
\label{T_rho}k T =\hbar c\left[\frac{45(1-\beta)}{\pi^2\mu m_\mathrm{H}\beta}\right]^{1/3}\rho^{1/3}.
\end{eqnarray}
Thus $P = \kappa \rho^\Gamma$, which implies that the state equation of the fluid is polytropic, therefore its internal
energy is proportional to the pressure $\epsilon=P/(\Gamma-1)$, and the
equation of motion of the fluid (\ref{torus_state_eq}) becomes
\begin{eqnarray}
\label{torus_state_eq1}\left({\rho+  \frac{\Gamma}{\Gamma-1}P}\right)a_\alpha=- \partial_\alpha P.
\end{eqnarray}
Substituting $\partial_\alpha P=\kappa \Gamma\rho^{\Gamma-1}\partial_\alpha \rho $ and $P = \kappa \rho^\Gamma$ into above equation, one obtains
\begin{eqnarray}
 \partial_\alpha \rho = -a_\alpha \left(\frac{\rho^{2-\Gamma}}{\kappa \Gamma}+\frac{\rho}{\Gamma-1}\right).
\end{eqnarray}
Introducing a new variable $\xi$ defined by $\xi=\ln(\Gamma-1+\kappa\Gamma\rho^{\Gamma-1})$, above equation
is simplified as
\begin{eqnarray}
\label{torus_equation}\partial_\alpha \xi = -a_\alpha,
\end{eqnarray}
which implies that the vector $\mathbf{n}=(a_r, a_\theta)$ in the $r$-$\theta$ plane can be regarded as the
normal vector of the contours of the density $\rho$. Thus if we use $\mathbf{t}=(dr, d\theta)$ to denote
the tangent vector of the contours, we have $\mathbf{n}\cdot\mathbf{t}=0$, or equivalently
\begin{eqnarray}
\label{s11}a_rdr+a_\theta d\theta = 0.
\end{eqnarray}
If we use $ds$ to denote the differential proper length of the tangent vector, we have
\begin{eqnarray}
\label{s22}ds^2= \mathbf{t}\cdot\mathbf{t}=g_{rr}dr^2+g_{\theta\theta}d\theta^2,
\end{eqnarray}
where $g_{rr}$ and $g_{\theta\theta}$ are the components of the Kerr metric, and $g_{rr}=\Sigma/\Delta$,
$g_{\theta\theta}=\Sigma$. Solving the equations (\ref{s11}) and (\ref{s22}) simultaneously, we get
a set of differential equations to describe the contours of density $\rho$
\begin{eqnarray}
&&\frac{dr}{ds}=\sqrt{\frac{\Delta}{\Sigma}}\frac{|a_\theta|}{\sqrt{a_\theta^2+\Delta a_r^2}},\\
&&\frac{d\theta}{ds}=-\sqrt{\frac{\Delta}{\Sigma}}\frac{|a_r|}{\sqrt{a_\theta^2+\Delta a_r^2}}.
\end{eqnarray}
If we introduce an auxiliary variable $\zeta$ defined by $d\zeta=\sqrt{\Delta/\Sigma}ds$, and substitute
equations (\ref{ar}) and (\ref{at}) into the above equations we get
\begin{eqnarray}
\label{torus_isobar11}\frac{dr}{d\zeta} = \frac{\psi_2}{\sqrt{\psi_2^2+\Delta \psi_1^2}},\\
\label{torus_isobar22}\frac{d\theta}{d\zeta} = \frac{-\psi_1}{\sqrt{\psi_2^2+\Delta \psi_1^2}},
\end{eqnarray}
which have the exactly same forms with equations (\ref{torus_isobar1}) and (\ref{torus_isobar2}), where
$\psi_1$ and $\psi_2$ are given by the equations (\ref{torus_contour1}) and (\ref{torus_contour2}).
With these equations, the distributions of the mass density $\rho$ of the torus now are readily to be computed
by evaluating the integral of $\xi$ from the
torus center $r=r_k$, $\rho = \rho_c$ to the location ($r$, $\theta$)
along a path C which is orthogonal to the density contours everywhere.
And the integral of $\xi$ is
\begin{eqnarray}
\xi = -\int_{C}a_r dr+a_\theta d\theta.
\end{eqnarray}
From the equations (\ref{p_rho}) and (\ref{T_rho})
one can get the total pressure and temperature distributions immediately with the given density $\rho$.

Knowing the structure model of the torus, the absorption and emission coefficients are now readily to be
specified, with which we can discuss the radiative transfer process inside the torus.
Using the above torus model, we shall give two examples of radiative transfer applications.

\begin{figure}[ht!]
\begin{center}
\includegraphics[width=0.8\textwidth]{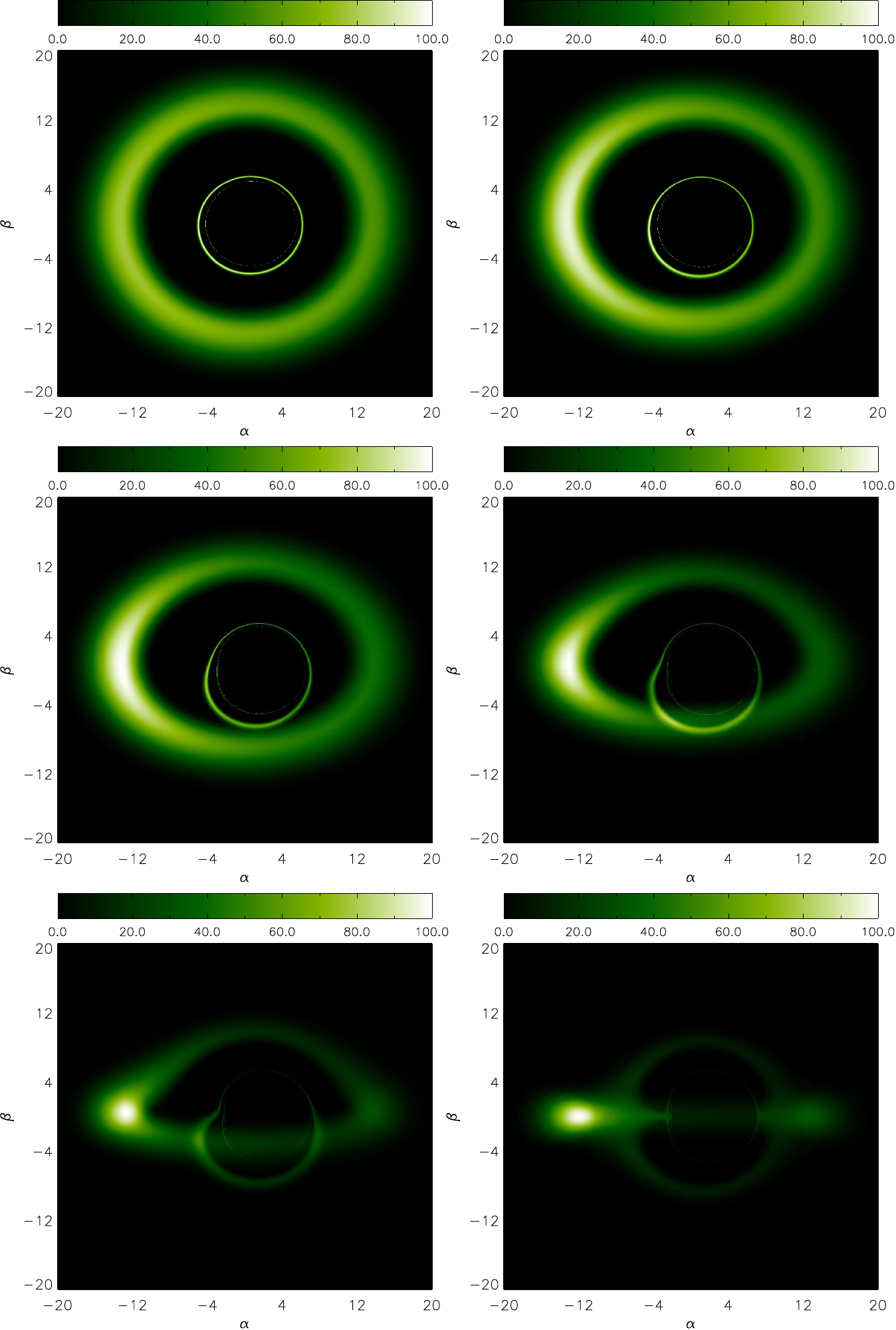}
\caption{\label{torusthin}
This figure shows the images of an optically thin and
radiation pressure dominated torus. The inclination angles of the observer
are $15^\circ$, $30^\circ$, $45^\circ$, $60^\circ$, $75^\circ$ and $90^\circ$ for
panels from left to right and top to bottom. The black hole spin $a$ is
$0.998$, and the ratio of gas pressure to total pressure $\beta$ is $2.87\times 10^{-8}$.
The torus parameters are $n = 0.21$, $r_k=12$ $r_g$. The brightness of each
pixel represents the observed intensity integrated along a geodesic ray at
a given frequency and has been normalized, and the maximum for
each panel is the same and equals to 100. $\alpha$ and $\beta$ are the impact parameters.}
\end{center}
\end{figure}

Firstly we consider a rather simple case, in which the torus is optically thin.
The emissivity is taken to be proportional to the mass density $\rho$, i.e., $j_{em}\propto\rho$,
and is independent on the frequency $\nu$. The absorption coefficient $\alpha_\nu$ is simply assumed to be zero.
The torus parameters are $n = 0.21$, $r_k=12$ $r_g$. The black hole
spin $a$ is $0.998$. The ratio of gas pressure to total pressure $\beta$ is $2.87\times 10^{-8}$.
In Figure \ref{torusthin} we draw the images of the torus, which is optically thin and radiation pressure dominated.
We see that the emissions mainly come from the central region of the torus, where the density is higher.
As the inclination angle of observer increases, the frequency shift of the emission
caused by the Doppler boosting becomes larger. In this figure the false color represents the
observed intensities of the emission, showing that the approaching side of the
torus is brighter than the receding side especially at higher inclination angles.

\begin{figure}[ht!]
\begin{center}
\includegraphics[width=0.8\textwidth]{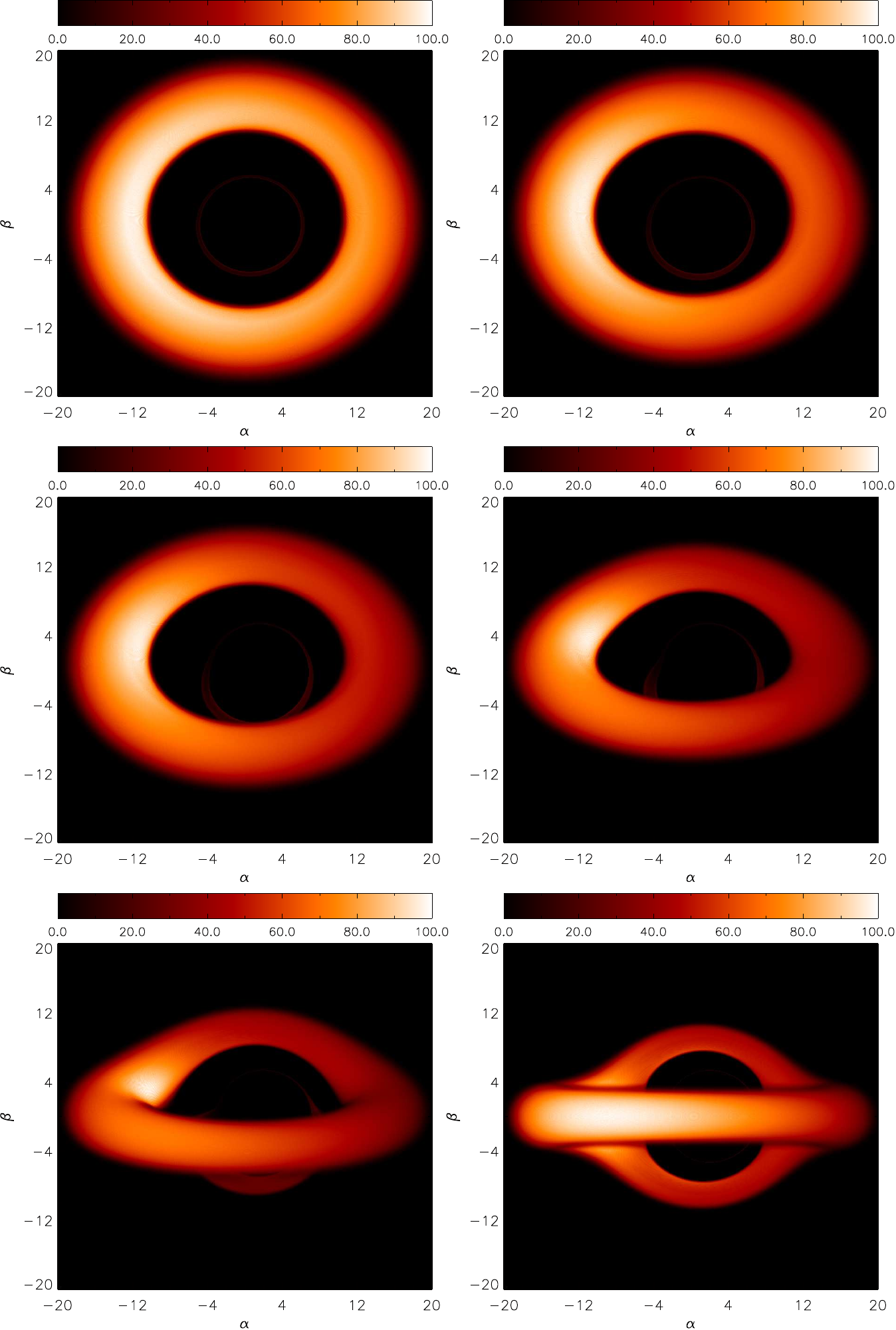}
\caption{\label{torus_thick}
This figure shows the images of an optically thick and
semi-opacity torus. The inclination angles
are $15^\circ$, $30^\circ$, $45^\circ$, $60^\circ$, $75^\circ$ and $90^\circ$ for
panels from left to right and top to bottom. The black hole spin $a$ is
$0.998$, and the ratio of gas pressure to total pressure $\beta$ is $2.87\times 10^{-8}$.
The torus parameters are $n = 0.21$, $r_k=12$ $r_g$. The brightness of each
pixel represents the observed intensity integrated over the entire spectrum. The
intensity has been normalized, and the maximum of each panel is the same and
equals to 100. $\alpha$ and $\beta$ are the impact parameters.}
\end{center}
\end{figure}

Secondly, we mimic a more realistic case, namely the thermal free-free emission and
absorbtion procedure, in which the torus is semi-opacity.
The emission and absorbtion coefficients of the torus for a photon at energy $E_0$ are given by \citep{Younsi2012}
\begin{eqnarray}
j(E_0)& =& \mathcal{K}\left(\frac{n_e}{cm^{-3}}\right)^2\left(\frac{E_0}{keV}\right)^{-1}\left(
\frac{\Theta}{keV}\right)^{-1/2}e^{-E_0/\Theta},\\
\alpha(E_0)& = &B_1\left(\frac{n_e}{cm^{-3}}\right)^2 \frac{\sigma_\mathrm{T}}{E_0^2},
\end{eqnarray}
where $\Theta = k_\mathrm{B}T$, $\mathcal{K}$ and $B_1$ are the normalization constants, $n_\mathrm{e}$ is
the electron number density and $n_\mathrm{e}=\rho/\mu m_\mathrm{H}$, $\sigma_\mathrm{T}$ is the
Thompson cross-section. The observed intensity images of the optically thick and semi-opacity torus are
plotted in Figure \ref{torus_thick}.
These images are quite different from
those of an optically thin torus. The emissivity now depends on the temperature,
which decreases towards to the outer surface of the torus, leading the limb darkening
phenomenon. When the rays are nearly tangential to the layers of the torus, they will
travel a longer distance and go through the outer, thus colder layers. While when
the rays are perpendicular to the layers of the torus, they
will travel a shorter distance and go through the inner therefore hotter layers.
Consequently, the observed intensity at lower inclination angles will be
much brighter than that at higher inclination angles \citep{Younsi2012}.

\section{Discussions and conclusions}
\label{discconc}
Following \cite{dexagol2009} we have presented a new public code named ynogk
for the fast calculating of null geodesics in a Kerr spacetime. The code is written
by Fortran 95, and composed by a couple of modules. In which the object-oriented
method has been used, which makes the addition of the code to one's own
readily.

In ynogk the B-L coordinates $r$ and $\mu$ have been expressed as analytical functions of the
parameter $p$. In these expressions, the Weierstrass' and Jacobi's elliptic function $\wp(z;g_2,g_3)$,
$\mathrm{sn}(z|k^2)$ and $\mathrm{cn(z|k^2)}$ are used, since the reductions to Weierstrass's
standard integrals are much easier, in which only one real root of the equations $R(r)=0$ and
$\Theta_\mu = 0$ is required. The B-L coordinates $t$, $\phi$ and the affine parameter $\sigma$
have been expressed as numerical functions of $p$. For a given $p$, the number of times of a photon
reaches the turning points both in radial and poloidal motions is uniquely determined and
needs not to be specified by the user.

Actually in addition to $p$, coordinates $r$, $\mu$ (or $\theta$) can also be
taken as the independent variables \citep{dexagol2009}.
The main reason of using $p$ is that one can
pay no attention to handle turning points,
which has been done by the inner routines of our code. This virtue is convenient
for a person who is not familiar with or has no interesting to
the details of the calculation of a geodesics
in the Kerr spacetime. Another reason is that
the value of $p$ which corresponds to the termination of the geodesic---either
at the infinity or the event horizon---is finite.
Thus it is easier to handle $p$ than $r$. In
our code $r$ and $\mu$ can also be taken as the independent variable. We provide
a routine named {\bf{geokerr}}, which can take $r$ or $\mu$ as the independent
variable. But the number of turning points should be prescribed.

With the expressions of all coordinates and affine parameter as functions of $p$,
the ray-tracing problem, which determines
the intersection of the ray with a target object, now becomes a root finding
problem. The function $f(p)$ that describes the
surface of the target object needs to be given by the user and the roots of equation
$f(p)=0$ correspond to the intersections. We provide a module named
{\bf{pem-finding}} to search the minimum root of this equation by
the Bisection or the Newton-Raphson method. In addition,
the observer-emitter problem can also be converted to a root finding problem,
which requires one to solve a set of nonlinear equations. A module named
{\bf{obs-emitter}} based on the Newton-Raphson method to solve these equations
is provided in our code. The routines in this module will return the solution,
provided the coordinates of the
emitter, $r_{em},\theta_{em}$ and $\phi_{em}$, are given.

We present a new set of formulae to compute the constants of motion $\lambda$ and $q$ from
initial conditions. These formulae can be regarded as the extensions of
\cite{cunnbard1973}. Our formulae are pervasive and can be used to handle
more sophisticated cases, in which the motion state and the finite distance
of the observer or the emitter with respect to the black hole are considered. One may find it is
convenient when dealing with problems in which the emitter has motion
and is closed to the vicinity of a black hole, e.g., the self-irradiation process
in the inner region of a disk.

The code has been tested extensively with various toy problems in the literature. The results
agree well with previous works. The comparisons with
geokerr of \cite{dexagol2009} also have been presented.

Finally we point out that the strategy discussed in this paper can be naturally extended to the calculation
of the timelike geodesics almost without any modification. Especially for the
timelike bounded orbits, in which the number of turning points both
in poloidal and radial coordinates can be arbitrary. The extension of this strategy
to calculate the timelike geodesics in a Kerr-Newmann spacetime has been done and
the results are under preparation.

\acknowledgments
\section*{Acknowledgments}
We acknowledge the financial supports from the National
Basic Research Program of China (973 Program 2009CB824800), the National Natural
Science Foundation of China 11163006, 11173054, and the
Policy Research Program of Chinese Academy of Sciences (KJCX2-YW-T24).
We also thank the anonymous referee for
very creative and helpful comments and suggestions, which have improved both our work
and the paper much.


\clearpage

\clearpage
\end{document}